\documentclass[12pt,preprint]{aastex}
\usepackage{aas_macros}

\def\lesssim{\mathrel{\hbox{\rlap{\hbox{\lower4pt\hbox{$\sim$}}}\hbox{$<$}}}}
\def\gtrsim{\mathrel{\hbox{\rlap{\hbox{\lower4pt\hbox{$\sim$}}}\hbox{$>$}}}}

\newcommand{\gapp}{\mbox{\raisebox{-0.3em}{$\stackrel{\textstyle >}{\sim}$}}}
\def\arcsec{$^{\prime\prime}\,$}
\def\arcmin{$^{\prime}\,$}
\def\deg{$^{\circ}$\,}

\def\gtrsim{\mathrel{\hbox{\rlap{\hbox{\lower4pt\hbox{$\sim$}}}\hbox{$>$}}}}
\newcommand{\ltappeq}{\raisebox{-0.6ex}{$\,\stackrel{\raisebox{-.2ex}{$\textstyle <$}}{\sim}\,$}}
\newcommand{\gtappeq}{\raisebox{-0.6ex}{$\,\stackrel{\raisebox{-.2ex}{$\textstyle >$}}{\sim}\,$}}

\def\spose#1{\hbox to 0pt{#1\hss}}
\def\lta{\mathrel{\spose{\lower 3pt\hbox{$\mathchar"218$}} \raise 2.0pt\hbox{$\mathchar"13C$}}}
\def\gta{\mathrel{\spose{\lower 3pt\hbox{$\mathchar"218$}} \raise 2.0pt\hbox{$\mathchar"13E$}}}

\def\sp{J2345-0449\,\,}

\shorttitle{Mega parsec radio jets in a spiral galaxy}
\shortauthors{Bagchi et al.}
\usepackage{color}

\begin{document}

\title{ Mega parsec relativistic jets launched from an accreting 
supermassive blackhole in an extreme spiral galaxy}

\author{Joydeep Bagchi\altaffilmark{1}, Vivek M.\altaffilmark{1}, Vinu Vikram\altaffilmark{2}, 
Ananda Hota\altaffilmark{3}, Biju K.G.\altaffilmark{4,7}, 
S. K. Sirothia\altaffilmark{5}, Raghunathan Srianand\altaffilmark{1}, 
Gopal-Krishna\altaffilmark{1,6},  Joe Jacob\altaffilmark{7}}
     
\altaffiltext{1}{The Inter-University Centre for Astronomy and Astrophysics (IUCAA),  Pune University Campus, Post Bag 4, Pune  411007, India; joydeep@iucaa.ernet.in}
\altaffiltext{2}{Department of Physics and Astronomy, University of Pennsylvania, Philadelphia 19104, USA}
\altaffiltext{3}{UM-DAE Centre for Excellence in Basic Sciences,  Vidhyanagari, Mumbai  400098, India}
\altaffiltext{4}{Department of Physics, W.M.O. Arts \& Science College, Post Office Muttil, North Kalpetta,
      Wayanad, India}
\altaffiltext{5}{National Centre for Radio Astrophysics (NCRA), TIFR, Pune University Campus, Post Bag 3,
Ganeshkhind, Pune 411 007, India}
\altaffiltext{6}{NASI Platinum Jubilee Senior Scientist}
\altaffiltext{7}{Department of Physics, Newman College, Thodupuzha 685 585, India}

\begin{abstract}

Radio galaxy phenomenon is directly connected to mass accreting, spinning supermassive black holes 
found in the active galactic nuclei (AGN). It is still unclear how the collimated jets of relativistic plasma on hundreds to thousands of kpc scale form, and why nearly always they are launched from the nuclei of bulge dominated elliptical galaxies and not flat spirals. Here we present the discovery of giant radio source J2345-0449 ($z=0.0755$), a clear and extremely rare counter example where relativistic jets are ejected from a luminous and massive spiral galaxy on scale of $\sim1.6$~Mpc, the largest known so far. Extreme physical properties observed for this bulgeless spiral host, such as its high optical and infra-red luminosity, large dynamical mass, rapid disk rotation, and episodic jet activity are possibly the results of its unusual formation history, which has also assembled, via gas accretion from a disk, its central black hole of mass $>2 \times 10^{8}$~M$_{\odot}$. The very high mid-IR luminosity of the galaxy suggests that it is actively forming stars and still building a massive disk. We argue that the launch of these powerful jets is facilitated by an advection dominated, magnetized accretion flow at low Eddington rate onto this unusually massive (for a bulgeless disk galaxy)  and possibly fast-spinning central black hole. Therefore, J2345-0449 is an extremely rare, unusual galactic system whose properties challenge the standard paradigms for black hole growth and formation of relativistic jets in disk galaxies. Thus,  it  provides fundamental insight into accretion disk -- relativistic jet coupling processes.

\end{abstract}



\section{INTRODUCTION}

It is now widely recognized that supermassive black holes (SMBH)
of masses $\sim 10^{6}$-$10^{10}$~$M_{\odot}$ lurk in the nuclei of almost all massive 
galaxies \citep[e.g.,][]{KR95,Magorrian98}. A symbiotic relationship between the
growth of the central black holes  and bulges of galaxies is
suggested by a remarkably tight correlation between the masses of the black holes and  the
galactic bulges \citep{Magorrian98,Gebhardt2000,Haring_Rix,Marconi_Hunt}. The most spectacular  manifestations of
massive black holes in AGN are the  powerful bi-polar relativistic jets,  that 
form twin-lobed giant radio galaxies on $10^{2}$-$10^{3}$ kilo parsec (kpc) scale, 
and the extremely luminous quasars. Gravitational accretion of matter onto the SMBH is believed to
be the `central engine' that powers  such sources \cite{Lynden_Bell69,Soltan82}  and their growth
 governed by accretion processes \cite{Soltan82,BBR}.
Understanding the physics of black hole formation and relativistic jets is a major focus of modern astrophysics.
Powerful radio jets on $\gtrsim$100~kpc scale are  nearly  always launched
from the nuclei of elliptical galaxies and not  spirals, and  the typical radio
luminosity of  spiral galaxies with AGN is about $10^{3}$-$10^{4}$ times feebler than ellipticals, making
them comparatively radio-quiet. 
Moreover, while both ellipticals and spirals  may  host radio-quiet quasars,  radio-loud quasars are never 
found in spiral galaxies, but only in ellipticals, and about $10\%$ of quasars are
radio loud at a given epoch, for reasons that  are still unknown \citep[e.g.,][]{Dunlop03}.   

The  physical origin of  radio-loud/radio-quiet dichotomy  and mechanisms  by which relativistic jets are 
launched from accretion disks around black holes (the disk-jet coupling) have long been 
the subject of intense investigations, yet the issue still remains  unresolved
 despite a wealth of observations. It is plausible 
that the prime difference between radio-loud and radio-quiet AGN is connected to the 
fundamental black hole properties; namely its mass, spin and the accretion rate. 
Any theoretical model must explain why it is so difficult for massive black holes in disk galaxies to
eject collimated radio jets extending to $10^{2}$-$10^{3}$~kpc distances.  The question whether 
the  black hole mass or spin and radio loudness of  AGN host are interconnected remains unanswered  and is 
still under close scrutiny.  
It was  suggested by   {Laor \cite{Laor}} that only AGNs with 
black hole masses  $\gapp 3 \times 10^{8}$~$M_{\odot}$ in elliptical galaxies
produce large scale jets,  in contrast to  smaller mass black holes found  in spirals which obviously lack
such jets. According to the `spin-paradigm',  powerful large scale jets  originate near  rapidly spinning  
accreting SMBHs \cite{WC95,Sikora07,TNM10,Dotti13} found in bulge dominated systems, 
and launched at relativistic speeds via the magneto-hydrodynamic (MHD) 
Blandford-Znajek (BZ) mechanism \cite{BZ77,MT82,PNS13}, although 
observational evidence for this conjecture is sparse \cite{Wang06,Doeleman12}. Alternatively, in  
the Blandford-Payne (BP) mechanism \cite{BP82},  jet power is extracted 
from  the rotation of the accretion disk itself, via the magnetic field threading it,  
without invoking a rapidly spinning black hole. However, in both the processes
the intensity and geometry of the magnetic field near the black hole horizon strongly influences the Poynting flux of
the emergent jet \cite{BH08}. This  has given rise to the  `magnetic flux paradigm' which posits \cite{SB13} 
that jet launching and collimation requires strong magnetic flux anchored 
to an  ion-supported torus of
optically thin, geometrically thick, extremely hot gas having poor radiative efficiency \cite{Rees82}. 
 Such hot, magnetic field saturated ion-tori may arise naturally around spinning black holes via  
 advection dominated accretion flows (ADAF)  at low Eddington rates (Eddington luminosity for a 
 black hole of mass $M$ is; ${{L}_{Edd}} =  1.26 \times 10^{38} M/{M_{\odot}}\, {\rm erg~s^{-1}}$),  particularly 
in elliptical galaxies showing large scale jets \cite{Narayan94,Narayan95}. On the other hand,  
for AGN in spirals  (narrow-line Seyfert~1s, optically selected quasars), 
the mass accretion rates are plausibly near Eddington,  forming  
optically thick, geometrically thin  accretion disks which radiate efficiently,  but
fail to create  large-scale collimated radio jets.   


The physical mechanism  linking the super massive black hole mass to the galactic bulge properties is  
little understood, while it is believed that an 
energetic  AGN feedback, either via the mechanical power of the radio jets,  or 
through a pressure driven wind, on the 
surrounding  medium, plays a major role in regulating the growth of the black hole and even the 
surrounding host galaxy. Important  details of how SMBH grow, and their energetic feedback processes 
are  still missing, but it is gradually becoming clear  that SMBH
are essential ingredients shaping the lives of galaxies across the cosmic time \citep[for review see][]{Hickox12}. 
Thus a detailed study of  active galaxies hosting massive black holes  assumes great importance; 
particularly finding any clear counter examples of jet launching under  extremely unusual circumstances, 
such as the case of 
radio galaxy \sp presented in this work, may play a key role in  informing  us 
what factors determine the relativistic jet formation and its role in galaxy evolution.

 Our paper is organized as follows; In Sec.~\ref{radio_prop} we introduce the  radio source \sp 
and highlight its main  radio and optical properties. In Sec.~\ref{Obs_analysis} we present
the radio,  optical and mid-IR observations of the source, describe the data analysis procedure in detail and 
  obtain the main results. In Sec.~\ref{results} we focus on the scientific content of the 
 results and discuss their  wider  astrophysical implications. 
Finally, in  Sec.~\ref{discussion} we present our main 
discussion and conclusions in light of the  results obtained above,   and  highlight  the possibilities of 
future multi wavelength observations for an in-depth understanding of this extraordinary galaxy.

We adopt a $\Lambda$CDM cosmology model with $H_0=70.5$~km~s$^{-1}$~Mpc$^{-1}$,
$\Omega_M=0.27$ and $\Omega_\Lambda=0.73$, which results in a scale  of 1.43
kpc~arcsec$^{-1}$  for a redshift $z=0.0755$. The radio spectral index $\alpha$ is 
defined as:  flux density$\,(S_{\nu})$ $\propto$ frequency$\,(\nu)^{-\alpha}$.

\section{ A JET-LAUNCHING MASSIVE SPIRAL GALAXY 2MASX~J23453268-0449256}
\label{radio_prop}
Our most important  finding  is the  
extraordinary galaxy 2MASX J23453268-0449256 at a redshift $z=0.0755$
(hereafter called J2345-0449), which is an extremely  rare  and clear example  of megaparsec (Mpc) scale
collimated  pair of relativistic plasma jets launched from  the nucleus of a  massive  spiral galaxy.
This radio galaxy was first mentioned by Machalski et al. \cite{Machalski07} as a
Mpc-scale object located at redshift $z=0.0757$, but they did not provide any  further hints about
its extraordinary nature. Our {\em Very Large Array} (VLA)  and 
{\em Giant Meterwave Radio Telescope} (GMRT)\footnote[1]{\url{http://gmrt.ncra.tifr.res.in/}} images in Figure~\ref{fig1}  show  that  the synchrotron  
radio emission arises from an
enormous  ($>$Mpc scale) bipolar structure with two pairs of  radio lobes, 
i.e. forming a  giant  `double-double' radio galaxy (DDRG) centered on the spiral host.
With two nearly aligned pairs of radio lobes sharing a common  AGN core,
DDRGs are the most compelling, rare examples of
recurring jet activity of SMBHs \cite{Schoenmakers_2000,Saikia_Jamrozy_2009}. 
Remarkably, the inner  and outer radio lobe pairs   extend over $\sim387.2$~kpc ($\sim4.52$\arcmin)  
and  $\sim1.63$~Mpc ($\sim19.1$\arcmin), respectively,
making \sp  the largest radio source  known till date, hosted  by a  spiral galaxy.
The VLA 6cm (4.8~GHz) image shows that inner  `active'  radio lobes  are  of edge-brightened
(Fanaroff-Riley class II or FRII \cite{FR}) morphology, being fed by the 
collimated  jets shot out from the central nucleus, which is typical  of
DDRGs \cite{Schoenmakers_2000,Saikia_Jamrozy_2009}. 
The GMRT low frequency  330~MHz image shows that the `aged' outer lobes are more diffuse, 
lacking prominent hot spots,  no longer being energized by   jets. 
The diffuse steep-spectrum radio emission near inner lobes is
better visible in the lower frequency image, tracing the  back flowing plasma of outer lobes.
The integrated flux density and radio luminosity at 1.4~GHz are $S_{1.4} = 180.60\pm20.0$~mJy and
$L_{1.4} = 2.5 (\pm 0.3)\times 10^{31}$ erg~s$^{-1}$~Hz$^{-1}$ respectively. Although moderate, this  luminosity is
 much above the  rough  divide between the radio-quiet and radio-loud galaxies which
 is usually taken at $L_{1.4} \sim 10^{30}$ erg~s$^{-1}$~Hz$^{-1}$. However,  despite a clear
  FRII morphology on large scales, the 1.4 GHz radio luminosity
 of \sp is close to  the luminosity break between the FRI and FRII radio galaxies. 
The integrated flux density and radio luminosity at 330 MHz are $S_{330} = 3.60\pm0.15$~Jy and 
$L_{330} = 5.0 (\pm 0.2)\times 10^{32}$ erg~s$^{-1}$~Hz$^{-1}$ respectively, implying a 
very steep spectral index  ${\alpha^{1.4}_{330}} \approx 2$ between 1.4 GHz and 330 MHz.  
Both the outer  lobes have a very steep integrated radio spectral index ${\alpha^{1.4}_{330}} \sim 2$.
We observe  a sharp  increase in ${\alpha^{1.4}_{330}}$ between the inner and outer lobes on both sides of
the flat spectrum core, as expected  due to   radiative ageing of radio plasma in  the episodic 
jet activity scenario  (Figure~\ref{fig2}).

The galaxy's favorable inclination  the  line-of-sight  ($i\,$$\simeq59^\circ$)
has enabled a direct view of the well-developed, bright spiral arms, as seen from  the
deep, high resolution  optical images taken
with the {\em MegaCam} camera on the Canada-France-Hawaii Telescope (CFHT) and
also  visible  in the {\em Sloan Digital Sky Survey}  (SDSS) 
multi-band images \cite{Ahn14} shown in Figures~\ref{fig3} 
and \ref{fig5}. The {\it Fermi} Gamma-ray telescope  observations have 
recently indicated the presence of  relativistic jets 
in  a handful of spiral galaxies hosting narrow-line Seyfert~1  (NLSy1) nuclei \cite{abdo09}.
While those  jets are also very rare,  they are not known to extend much beyond the galactic 
scale ($\lesssim 10$~kpc) and lack the excellent collimation witnessed in the  
large-scale FRII jets \cite{Doi,Morganti}. It has been pointed out 
\cite{Foschini11} that the Blandford-Znajek mechanism fails to explain the jet power of these jets. 
Thus, the nature of radio emission in NLSy1s  and  their failure to produce large-scale radio jets   
has  remained unexplained. 
Despite  decades of extensive observations, 
only two previous reports exist of a disk galaxy ejecting large scale ($>$100~kpc)  bipolar radio jets:
the radio source J031552-190644  found in the galaxy cluster Abell 428 \cite{Ledlow_Owen_keel},
and a recently reported Mpc-scale episodic radio source  known as {\em Speca} \cite{Hota_2011}.
In both  these exceptionally rare  objects,   the  galactic disk is viewed 
 nearly edge-on, precluding a clear view of  the putative
spiral arms, and thus the evidence is still indirect. The present giant radio source \sp is not only the
most unambiguous example of this extremely rare class, but its spiral host galaxy also displays
a unique combination of several other remarkable  properties  (as discussed below). 
Extremely unusual  objects like \sp 
strongly challenge the conventional ideas of black hole growth and radio jet formation in 
galactic nuclei, thus, they are of profound  interest for  models of the central engines 
of radio galaxies and quasars.

We used SDSS-III images \cite{Ahn14} to explore  the  large scale galactic
environment in which  the spiral galaxy \sp  is located. The SDSS 
wide-field  color image in Figure~\ref{fig3}  and mid-IR data (Table~\ref{tab:infrared})
reveals that
\sp  is an unusually bright ($m_{r} = 14.40$, $M_{r} = -23.26$ 
in {\it r} band, $m_{12} = 9.00$, $M_{12} = -28.66$  in  $12 \mu$m  band) 
 isolated spiral galaxy   located in a sparsely 
populated galactic environment devoid of  bright L$^{*}$ galaxies, 
and clearly not at the center of a compact  group
or a rich galaxy cluster. The next brightest galaxy neighbour,  which is
located $\sim$460 kpc to the south east,  is more than two magnitudes fainter. The second and
third brightest galactic neighbours are \gtappeq2.5 magnitudes fainter and located $\sim$240 and 480 kpc away
respectively (Figure~\ref{fig3}). Interestingly, about 33 arcmin ($\sim$2.8 Mpc)
away from \sp (redshift $z = 0.07556$) is located a massive galaxy cluster
RBS~2042 (redshift $z = 0.07860$) which is strongly detected in the ROSAT All-sky Bright Survey
\cite{RASS,Schwope}.
Association of \sp with this cluster is not immediately obvious, but it could be
a peripheral  member of the  massive  cluster. 

\section{OBSERVATIONS AND DATA ANALYSIS}
\label{Obs_analysis}
\subsection{\bf GMRT observations \,\,}
The 325 MHz  data were obtained from the archives of  GMRT.
The observations were taken on  2008 November 25,  with the total on source integration time  
of $5.8$ hrs. The observations were taken at the frequency 325 MHz  with a 32 MHz band-width. 
The primary flux density calibrator  was 3C48
having flux density of 43.425 Jy at 325 MHz on the  VLA (1999) Perley-Taylor scale.
The phase calibrator  was  J2340+135 having a flux density  $7.74 $ Jy at 325 MHz.
The data reduction was done mainly using {\tt AIPS++}\footnote[2]{\url{http://aips2.nrao.edu/docs/aips++.html}}.
After applying bandpass corrections on the phase calibrator, gain and phase variations 
were estimated, and the flux density,
bandpass, gain and phase calibration  parameters from primary and phase calibrator were applied
on the target field. Bad quality data were flagged at various stages. The data for
antennas with high errors in the antenna-based solutions were examined and flagged over certain
time ranges. Some baselines were flagged based on closure errors on the bandpass calibrator.
Channel and time-based flagging of data points corrupted by radio frequency interference (RFI)
was done using a median filter with a $6\sigma$ threshold. Residual errors above $5\sigma$
were also flagged after a few rounds of imaging and self-calibration.
The system temperature ($T_{sys}$) was found to vary with antenna, the ambient temperature
and elevation. In the absence of
regular $T_{sys}$ measurements for GMRT antennas, this correction was estimated
from the residuals of corrected data with respect to the model data. The corrections
were then applied to the data. The final image  was made after several
rounds of phase self calibration, and one round of amplitude self-calibration,
where the data were normalized by the median gain for all the data. On cleaned map (Figure~\ref{fig1})
with FWHM beam $15.22^{\prime\prime} \times 11.35^{\prime\prime}$ at a position angle of $62.3^\circ$,
the root-mean-square noise level near the source is 1.1 mJy/beam.
On highest resolution map (not shown) of  FWHM beam $10.08^{\prime\prime} \times 7.53^{\prime\prime}$,
the rms noise level was 0.23 mJy/beam.

\subsection{\bf VLA observations   \,\,}
\begin{figure*}
\begin{center}

\includegraphics[scale=0.4, angle=0]{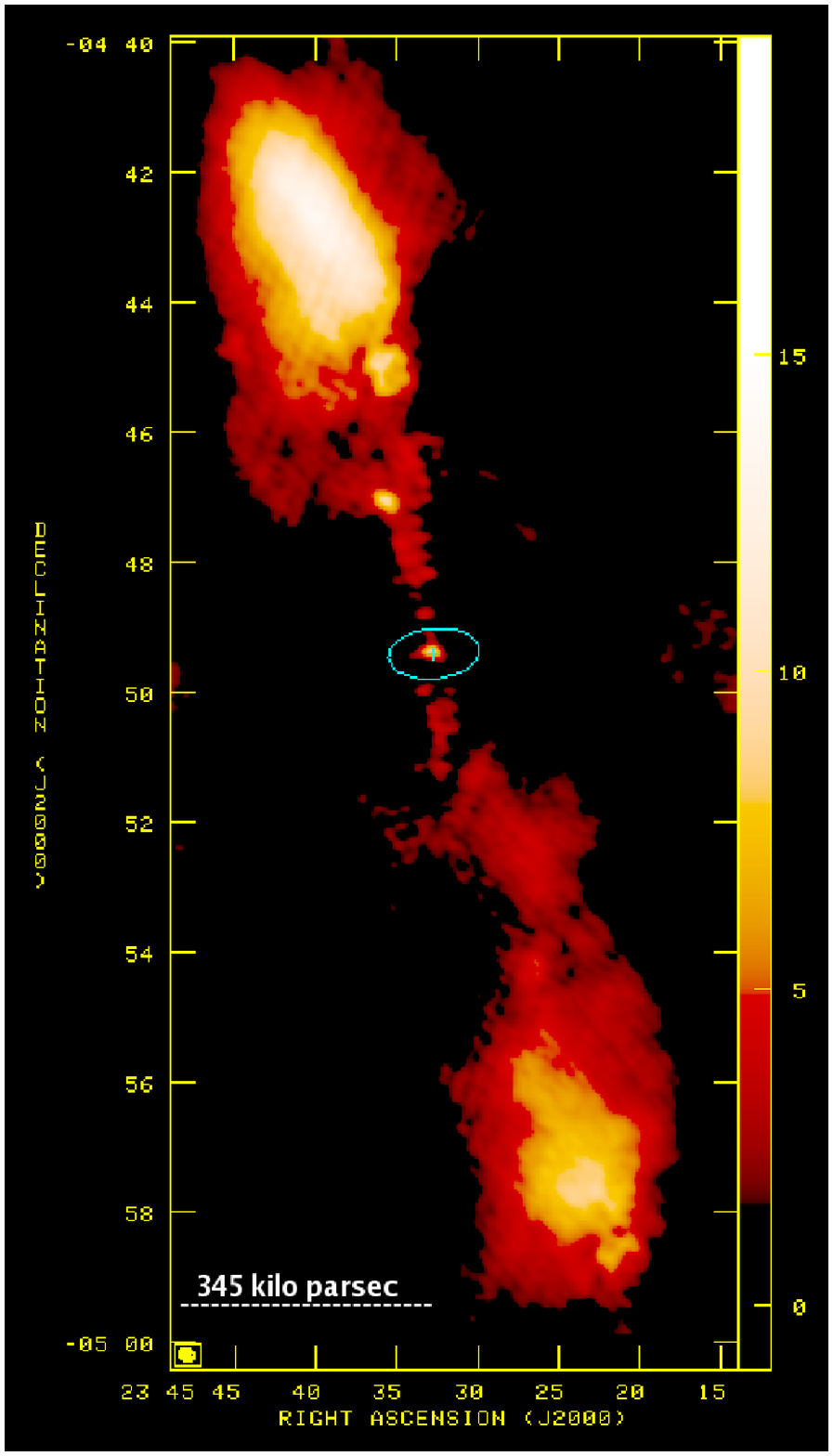}
\includegraphics[scale=0.4, angle=0]{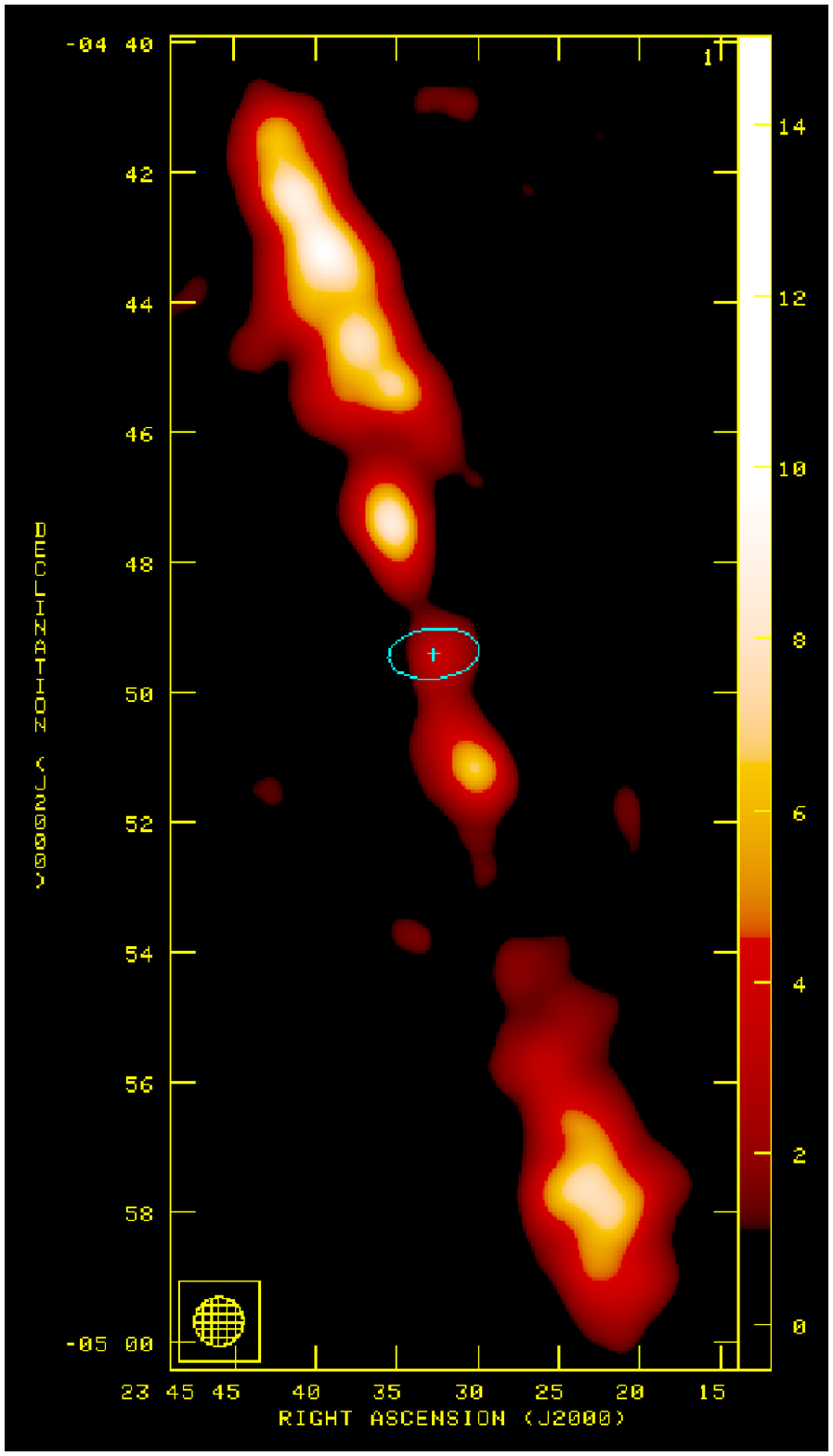}
\includegraphics[scale=0.28, angle=0]{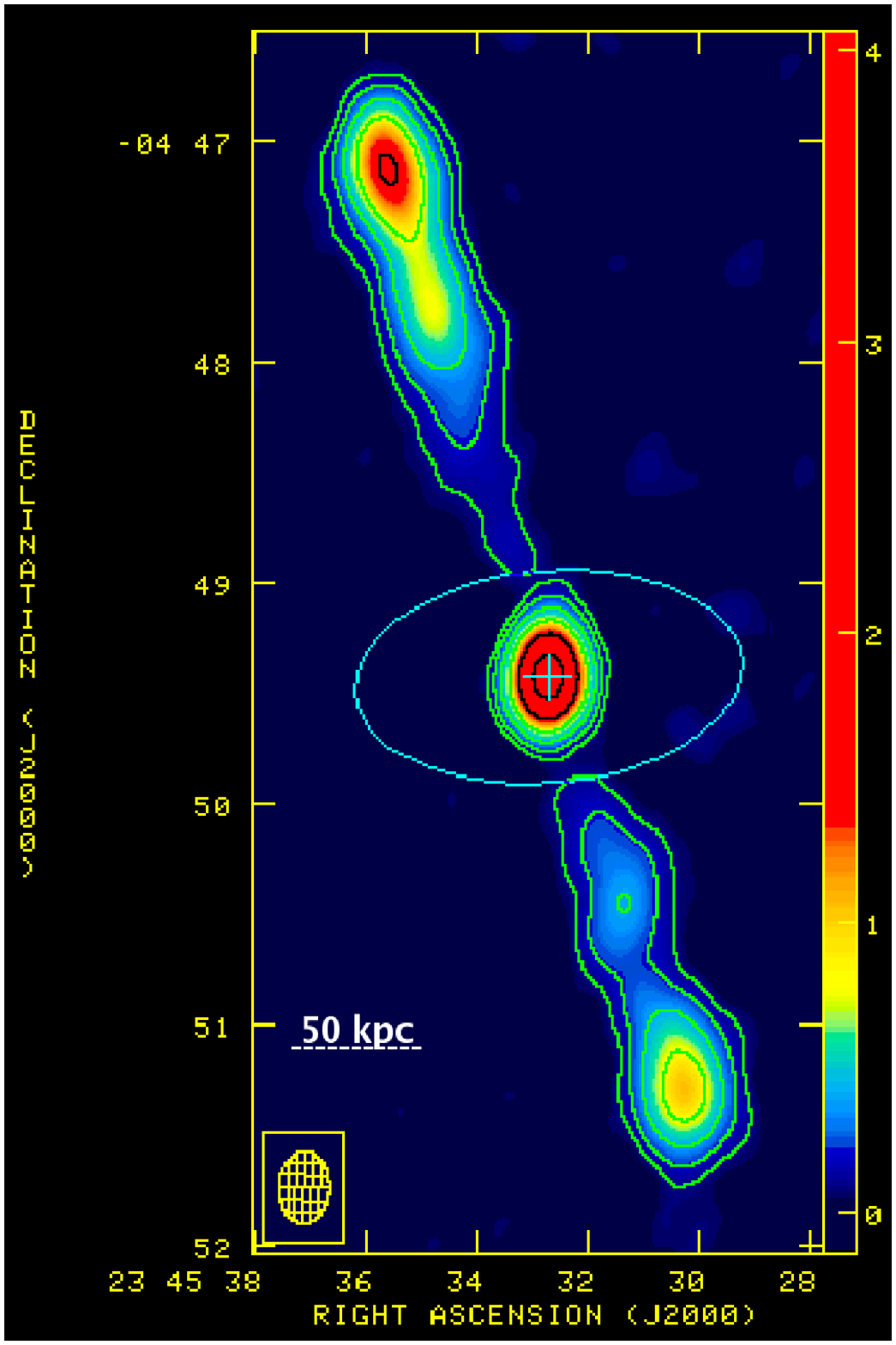}

\end{center}
\caption[]{{\bf The radio images of galaxy \sp.} \footnotesize (left) The 325 MHz image
taken with GMRT (beam FWHM: $15.22^{\prime\prime} \times 11.35^{\prime\prime}$ at $62.3^\circ$).
(right)  VLA 1.4 GHz image, taken from the NRAO VLA Sky Survey (NVSS), shown on the same scale
(beam FWHM: $45^{\prime\prime} \times 45^{\prime\prime}$). The diffuse emission  near
inner lobes is better visible at  lower frequency, tracing the
back-flowing   steep-spectrum plasma of the  outer  lobes. An outline of  the 
host spiral galaxy is drawn in blue, scaled to
about thrice the optical disk size for clarity. 
(bottom)  VLA 4.8~GHz image of the inner radio-double, showing the AGN core, bipolar radio-jets and the FR-II lobes
(beam FWHM: $19.8^{\prime\prime} \times 13.3^{\prime\prime}$ at  $178.5^\circ$). Contour levels
are: (-0.1, 0.1, 0.2, 0.4, 0.8, 1.6 and 3.2 mJy/beam). The spatial resolution (beam) is shown by
a yellow ellipse within a box in  each panel. The color wedges show the flux density levels in (mJy/beam). 
The bright, unresolved  radio core at 4.8 GHz coincides within 0.15 arcsec with the nucleus of
the spiral galaxy  at right ascension: $23h\ 45m\ 32.71s$,  
declination: $-04$\deg\ $49$\arcmin\ $25.32$\arcsec (J2000),
thus firmly establishing it to be the optical host of the radio galaxy.
}
\label{fig1}
\end{figure*}

\begin{figure*}
\begin{center}
\includegraphics[scale=0.46,angle=-90]{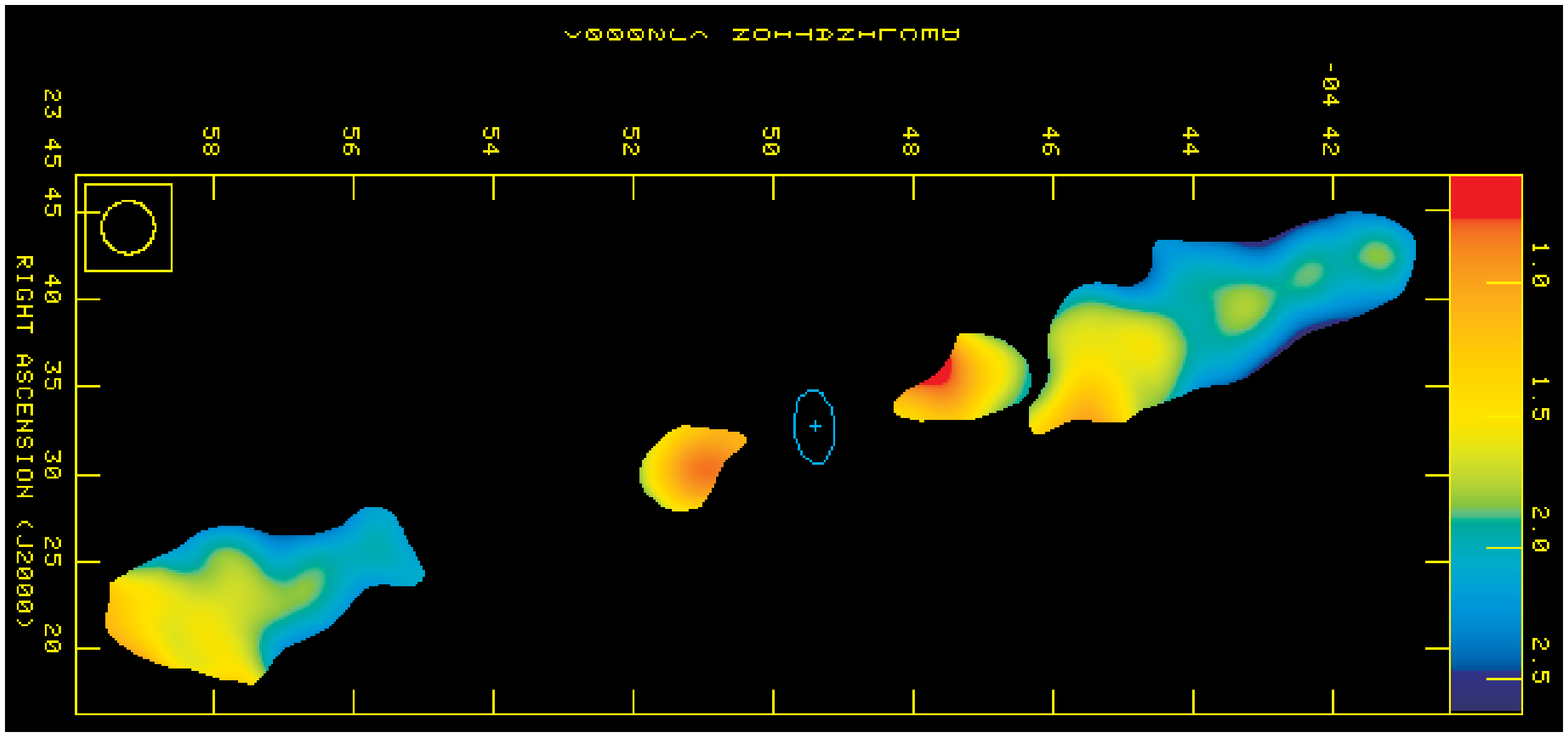}
\includegraphics[scale=0.5,angle=-90]{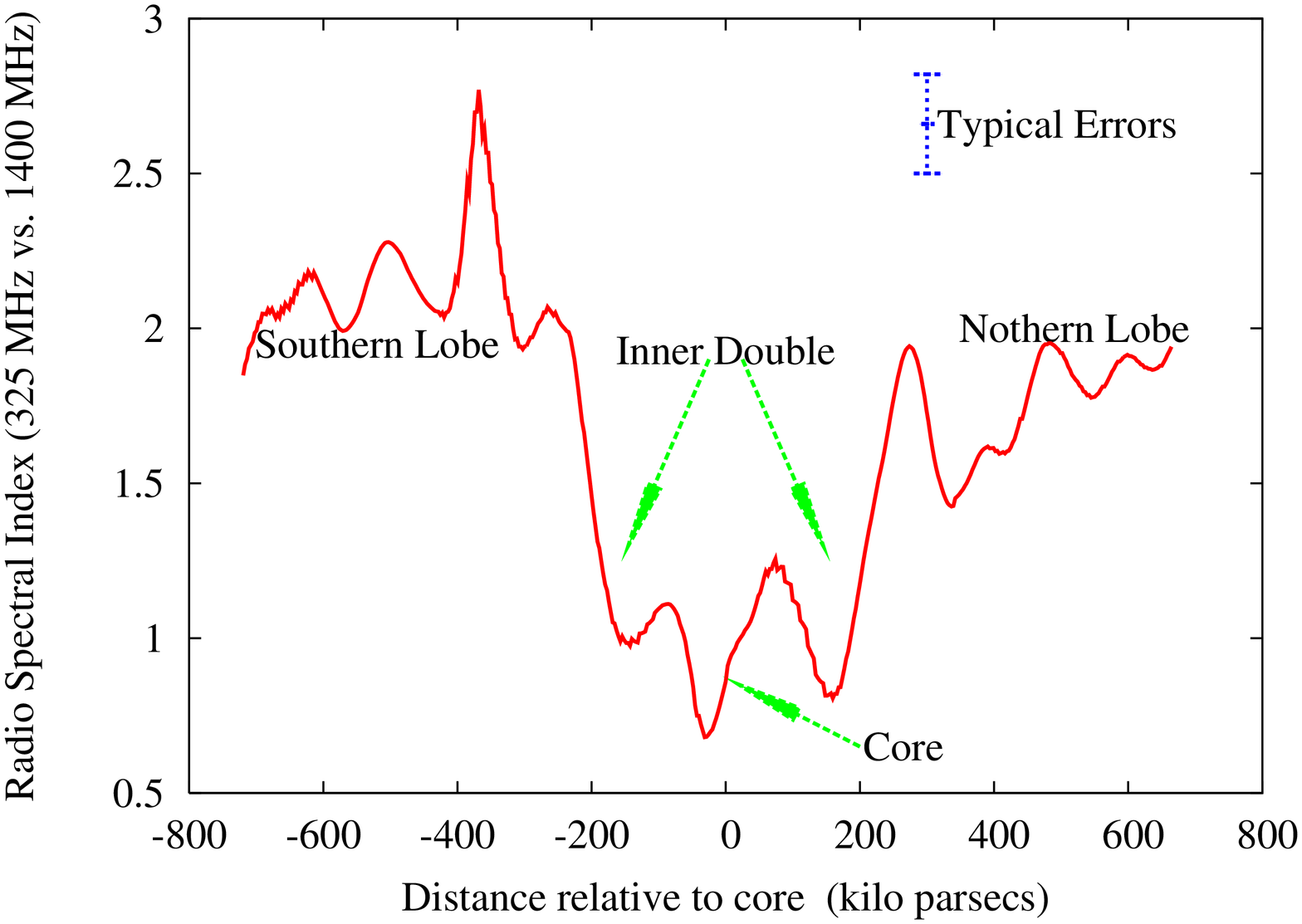}
\caption[]{  (Top) The  spectral index image (rotated by 90 degrees), and (Bottom) the 1D spectral
index profile, drawn along the brightness ridge of the source \sp between 
325 MHz (GMRT) and 1400 MHz  (VLA/NVSS) frequency is shown. In the spectral index map
 all pixels having flux density below 2 mJy/beam at 1400 MHz ($ \sim 4 \sigma$) and  15 mJy/beam at
  325 MHz ($ \sim 5 \sigma$)  have been blanked to prevent spurious structures from appearing in the map.
An outline of the optical galaxy is shown by a blue ellipse, and the
yellow circle within a box shows the $45^{\prime\prime}$$\times$$45^{\prime\prime}$ beam.
The spectral index $\alpha$ is defined as:  flux density $(S_{\nu})$ $\propto$ frequency$(\nu)^{-\alpha}$.
Typical error ($1~\sigma$) in the estimation of  spectral index is shown at the top right corner of 
the 1D profile figure. The positions of radio core, the inner  double lobes and the
outer lobes of the radio source  have been marked. Note a sharp increase in the spectral index 
between the inner and outer radio lobe pairs.
}
\label{fig2}
\end{center}
\end{figure*}

\begin{figure*}
\begin{center}
\includegraphics[scale=0.6]{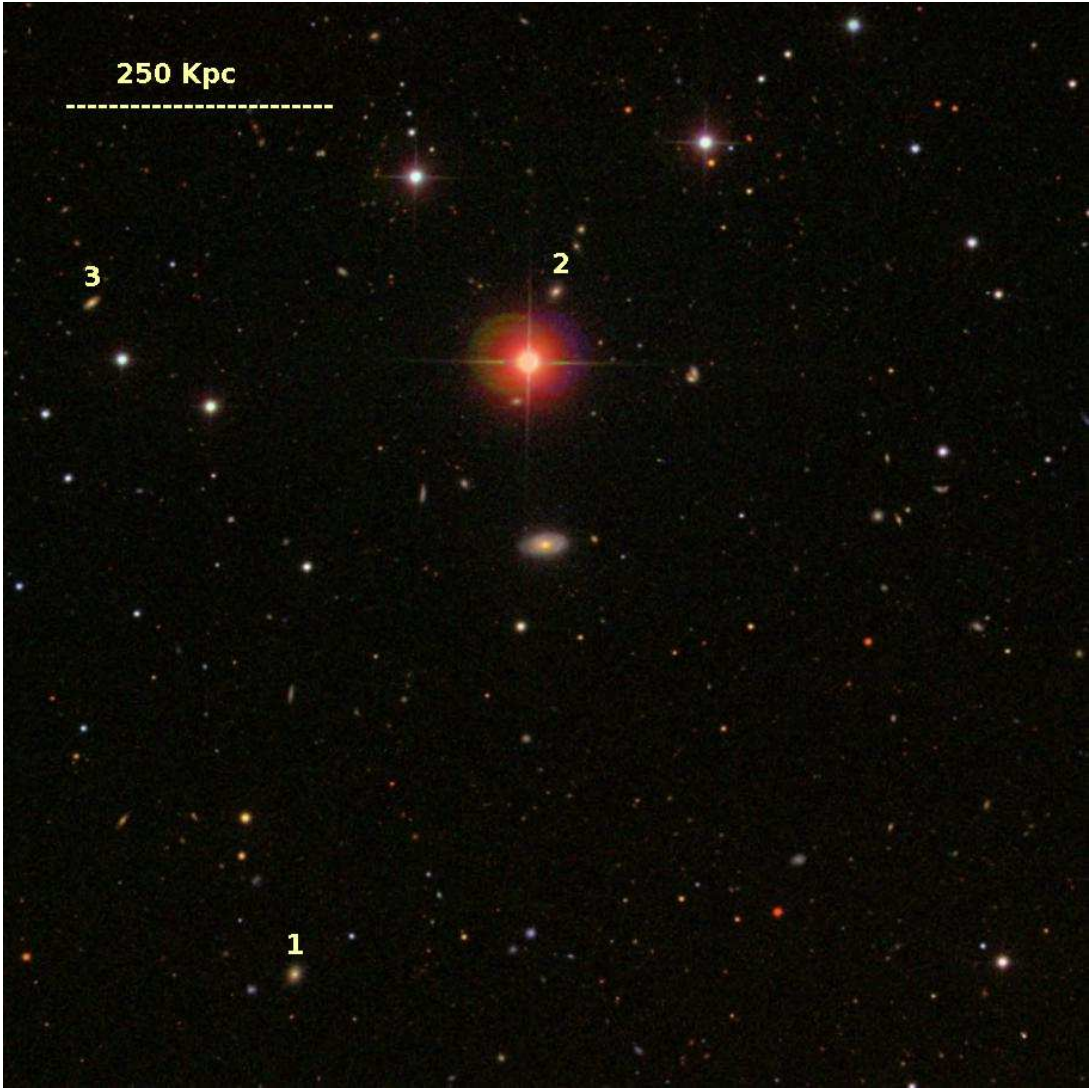}
\vspace{0.6cm}
\caption[]{\small The large scale galactic environment around J2345-0449.
The color image taken from
the {\em Sloan Digital Sky Survey-III},
shows a field of view of  $\sim 1\, {\rm Mpc} \times 1\, {\rm Mpc} $
centered on J2345-0449. The three nearest and brightest galaxies are marked in the descending order of
their brightness in {\it r}-band;  galaxy no.~1 ($m_{r} = 16.50$, $z_{\rm ph} = 0.0697$),
galaxy no.~2 ($m_{r} = 16.87$, $z_{\rm ph} = 0.0615$) and
galaxy no.~3 ($m_{r} = 17.07$, $z_{\rm ph} = 0.0798$). Here $z_{\rm ph}$ denotes  the photometric redshift obtained 
from SDSS. The magnitude of spiral galaxy \sp is $m_{r} = 14.40$  in {\it r} with  spectroscopic redshift 0.07556.
All magnitudes are galactic extinction corrected and taken from SDSS.
}
\label{fig3}
\end{center}
\end{figure*}

Very Large Array  (VLA)   C-band (6cm wavelength) data for
 \sp was obtained from the VLA archives. The observations were
taken on 2008 June 27, in D-configuration,  at the 4.88 GHz and 4.83 GHz centre frequencies,
and using two IF pairs of 50 MHz band-width each.  The primary flux density calibrator  was 3C48
having flux density of  5.517 Jy and  5.570 Jy for the two IFs on the  VLA (2010) flux
density scale. The total integration time at the radio galaxy's  core position
 was about 40~minutes and VLA  calibrator source J2358-103  was periodically
  observed for phase calibration.  For data reduction we used standard routines
    available in the {\em Astronomical Image Processing System} (AIPS) software.
     While calibrating,   bad quality data were flagged at various stages.
      After initial amplitude and phase calibration the data of two IFs were combined. Further data editing and
       final imaging using a visibility
        weighting scheme (ROBUST = 2.5)  in the imaging+deconvolution program IMAGR  resulted in good
	 quality radio maps  (Figure~\ref{fig1}). After deconvolution, the final clean
	  map was restored with an elliptical  Gaussian beam of FWHM
	   $19.8^{\prime\prime} \times 13.3^{\prime\prime}$ at  position angle
	    $178.5^\circ$. The root-mean-square noise on the  clean map, measured near the
	     centre was about $20 \mu$Jy/beam. The compact  AGN core  was strongly detected
	      with flux density $4.23 \pm 0.10$ mJy, and  peak located at 
	      (positional error  $\pm 0.15$ arcsec) right ascension: $23h\ 45m\ 32.71s$ 
	      and declination: $-04$\deg\ $49$\arcmin\ $25.18$\arcsec. The AGN radio  core  coincides
	        within errors with the optical nucleus of the spiral galaxy, measured on the CFHT  image at
		 right ascension: $23h\ 45m\ 32.71s$ and declination: $-04$\deg\ $49$\arcmin\ $25.32$\arcsec. The
		 positional coincidence  firmly establish the spiral  as  
		 the optical host of this giant radio galaxy (Figure~\ref{fig1}).

\subsection{\bf Evidence for interruption of the AGN jet activity\,\,}

Figure~\ref{fig2} shows the spectral index map and a 1D spectral index profile computed 
between GMRT 325 MHz and NVSS 1400 MHz,
after smoothing the 325 MHz map to the resolution of the latter, 
 i.e. $45^{\prime\prime} \times 45^{\prime\prime}$ beam.  In the spectral index map
 all pixels having flux density below 2 mJy/beam at 1400 MHz ($ \sim 4 \sigma$) and  15 mJy/beam at  
 325 MHz ($ \sim 5 \sigma$)  have been blanked to prevent spurious structures from appearing in the map.
The 1D profile is drawn along the ridge line of the source defined by the
brightness peaks seen on the full resolution 325 MHz GMRT map (Figure~\ref{fig1}). 
The most conspicuous aspect of the 
profile is the sharp  increase  in spectral index seen immediately
beyond the brightness peaks identified here as  the inner double. The inner `active'  lobes 
have relatively flatter spectra and are  fed  by the radio jets,  showing edge-brightened
FRII morphology,  consistent with their
interpretation as recently formed sources still being energized by the jets.
This behaviour is fully consistent with the canonical interpretation of such double-double radio sources
where the outer double lobes are viewed as the `relics' of a previous episode of nuclear
activity \cite{Schoenmakers_2000,Saikia_Jamrozy_2009}. 

\subsection{\bf Optical Spectroscopic Observations and Data Analysis\,\,}

Optical long-slit spectroscopic data  were taken
 during  2011 November 20-22  with  the 2m telescope
at the IUCAA Girawali Observatory (IGO). The spectra were obtained
using the IUCAA Faint Object Spectrograph and Camera (IFOSC) 
\footnote[3]{\url{http://www.iucaa.in/resources.html}}.
IFOSC employs an  EEV 2K$\times$2K, thinned, back-illuminated
CCD with 13.5$\mu$m pixels. The spatial sampling scale at the
detector is 44$\mu$m per arcsecond giving a field of view of about 10.5
arcminute on the side. We used two
IFOSC grisms;  grism No.7 and grism No.8  in combination with a 1.5-arcsec
slit, yielding a  wavelength coverage of 3800-–6840 \AA\,
and 5800–-8350 \AA\,  and spectral resolutions of 300 km/s and
240 km/s respectively. The spectra were taken with the slits aligned both   
along the major and minor axes of the galaxy.
In each case, we took 4 exposures (4$\times$ 45 min) using the
grism No.8 and 2 exposures (2$\times$ 45 min) using grism No.7,
respectively.  We carried out the bias and flat-field corrections to all the frames. 
Dark correction was not required as the CCD was liquid nitrogen cooled and the dark current was 
negligible for 45 min exposures. The cosmic ray hits were removed by the 
IRAF task `crmedian' in the Cosmic Ray Removal Utility Package (CRUTIL).
Standard stars were observed during the same nights by placing them at different slit locations. 
Wavelength calibration was done using the standard Helium-Neon lamp spectra.

\begin{figure*}
\begin{center}
\includegraphics[scale=0.6,angle=-90]{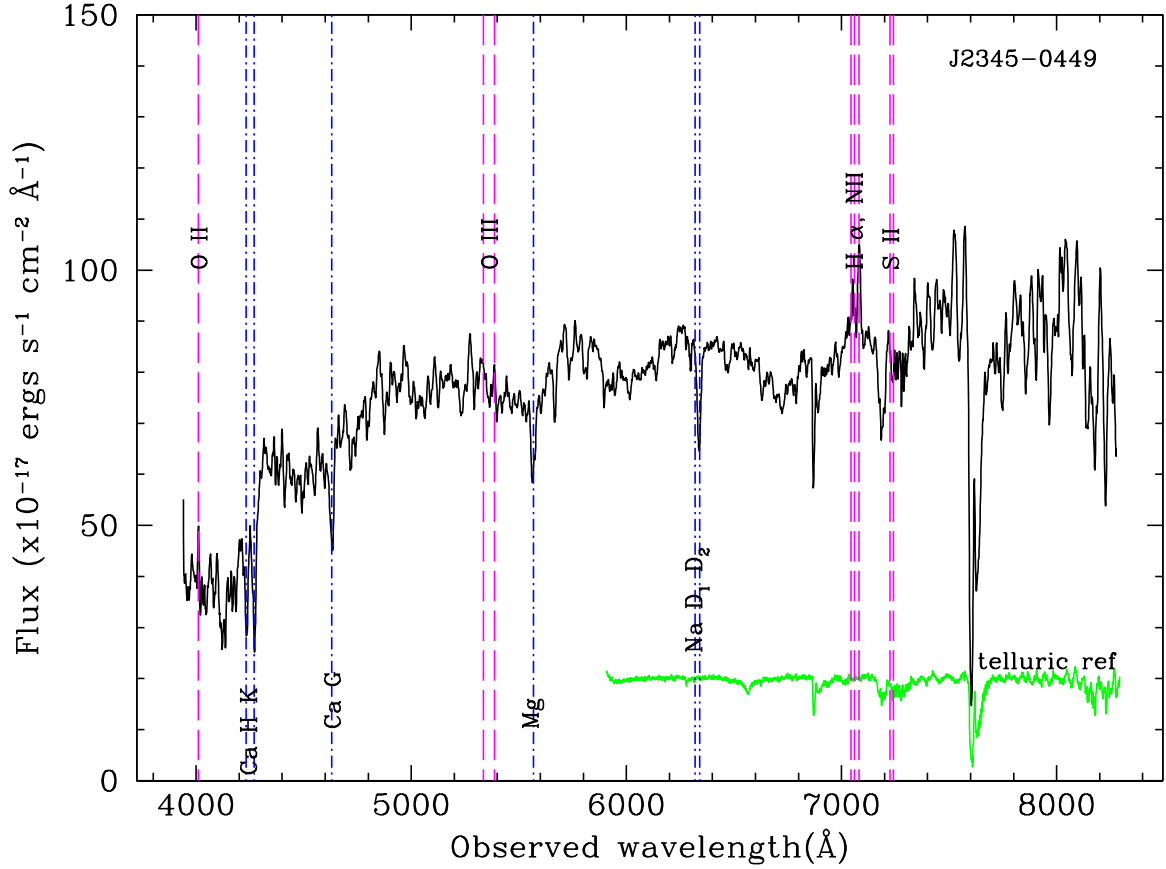}
\vspace{0.6cm}
\caption[]{\small Major axis spectrum taken with IFOSC on IUCAA 2m telescope, centered on the
nucleus. The absorption lines are marked by 
blue dot-dashed lines and the expected positions of the emission lines are marked with magenta 
color dashed lines. The red edge of the spectrum is contaminated by Telluric features. A Telluric reference 
spectrum is shown in green color at the bottom.   }
\label{ifosc_spect}
\end{center}
\end{figure*}

One dimensional spectra were extracted  using the {\em doslit} task in the IRAF software. We used
the variance-weighted extraction method as compared to the normal one. Air-to-vacuum conversion
was applied before combining the spectra using 1/$\sigma^2$ weighting in each pixel after scaling
the overall individual spectrum to a common flux level within a sliding window. The extraction was carried 
out both with and without cosmic ray removal. Fringing poses a problem for the grism No.8 (red region) 
data and we  removed the fringes by  subtracting two 2D science exposures for which the object was placed 
at two different locations along the slit.  From the absorption and emission features in the spectrum (shown in
Fig.~\ref{ifosc_spect}),  redshift of the galaxy was determined to be $\rm z = 0.0755\pm0.0005$.  We 
compared our spectrum with the 6dF spectrum given by Jones et al. \cite{Jones09}. The 
features in both the spectra  match closely and our redshift estimate is 
very close to the reported 6dF redshift $\rm z = 0.075566\pm 0.000150$.

As the galaxy is spatially well resolved, we also performed the 2D spectral extraction around
the H${\alpha}$ line detected in the grism No.8 spectra.  For spectral curvature corrections, the spectra 
of a standard star (Feige 110)  were taken at different spatial positions along the slit during the scheduled nights
and nearby nights. The same standard star was used for the flux calibration. 
We created a superimposed frame by joining the
different standard star spectra to the target spectra, retaining the original spatial position of each
spectra. The data were then corrected for the geometrical distortions 
using the IRAF tasks {\em fitcoords} and {\em transform}.
{\em Fitcoords} provides the transformation of slit position and wavelength as a function of (x,y) 
pixel number on the image. {\em Transform} applies this transformations to the data to correct the distortions. 
The resulting frame is used for the spectral extraction.

\subsection{\bf  Deriving the Galaxy Rotation Curve\,\,}
\label{rotcurv}
\begin{figure*}
\begin{center}
\includegraphics[scale=0.37,angle=0]{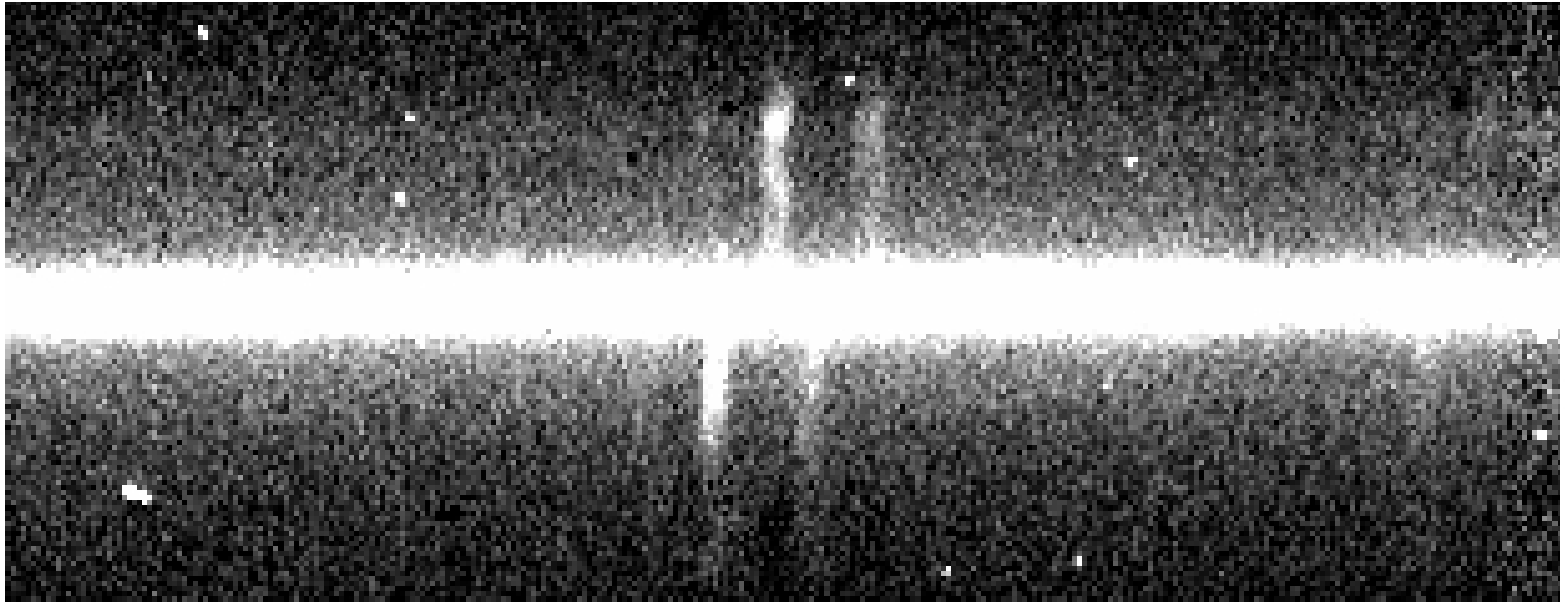}
\includegraphics[scale=0.35,angle=0]{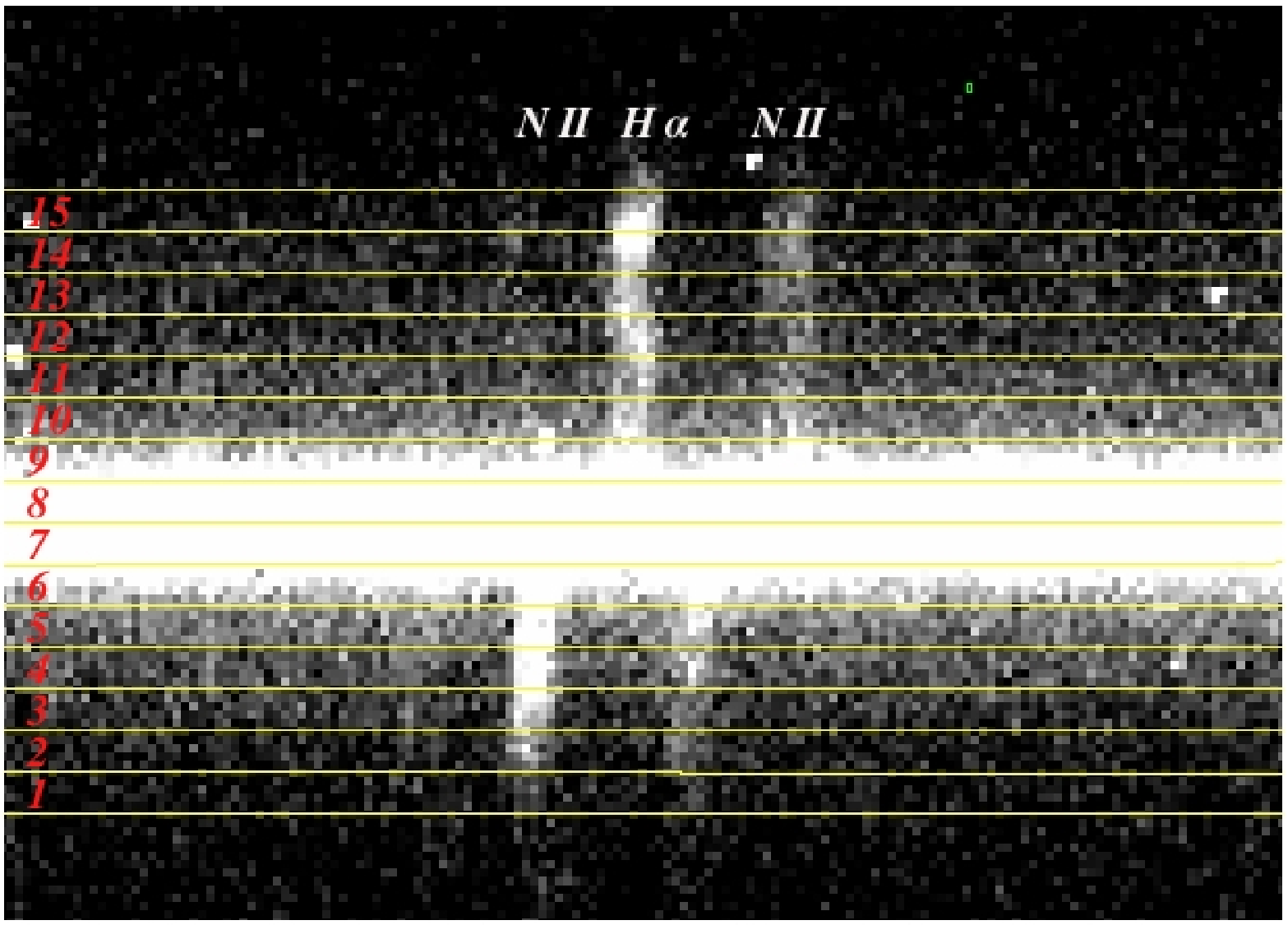}
\includegraphics[scale=0.55,angle=90]{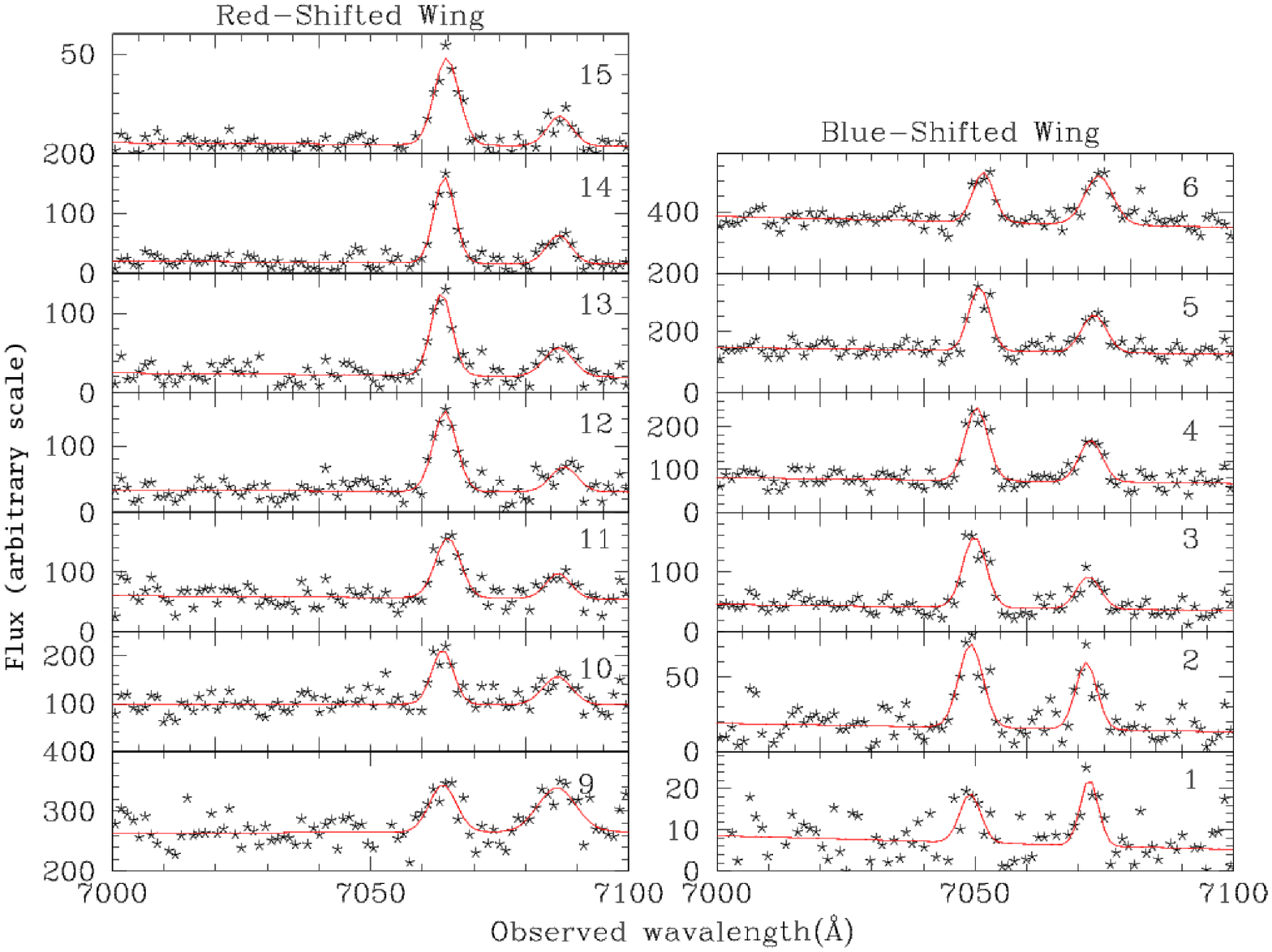}
\vspace{0.6cm}
\caption[]{\small Upper  panels: 2-D images showing the
long-slit spectrum near the H$\alpha$ and [NII] emission lines of galaxy \sp.
Locations of different sub-apertures (numbers 1 to 15) used to derive
the rotation curve of galaxy are  marked
in the zoomed right hand side image. Wavelength direction is horizontal, increasing to the right
and the spatial direction is vertical. A sharply rising flat rotation curve is clearly evident in
the shape of both  H$\alpha$ and [NII] lines marked here.
Lower panels: 1-D spectra extracted at different sub-apertures as
indicated in the above figure. Gaussian fits to H$\alpha$ and [NII] lines are shown for the blue and
red shifted wings of the spectra.  The flux scale is arbitrary. No  significant H$\alpha$ emission was detected in the
central apertures numbered 7 and 8. The parameters derived from the
Gaussian fits  are used to obtain the
rotation curve (Figure~\ref{fig5}),  and the
radial dependence of star formation rate and metallicity (Figure~\ref{fig6}).
}
\label{fig4}
\end{center}
\end{figure*}

\begin{figure*}
\begin{center}
\includegraphics[scale=0.65, angle=0]{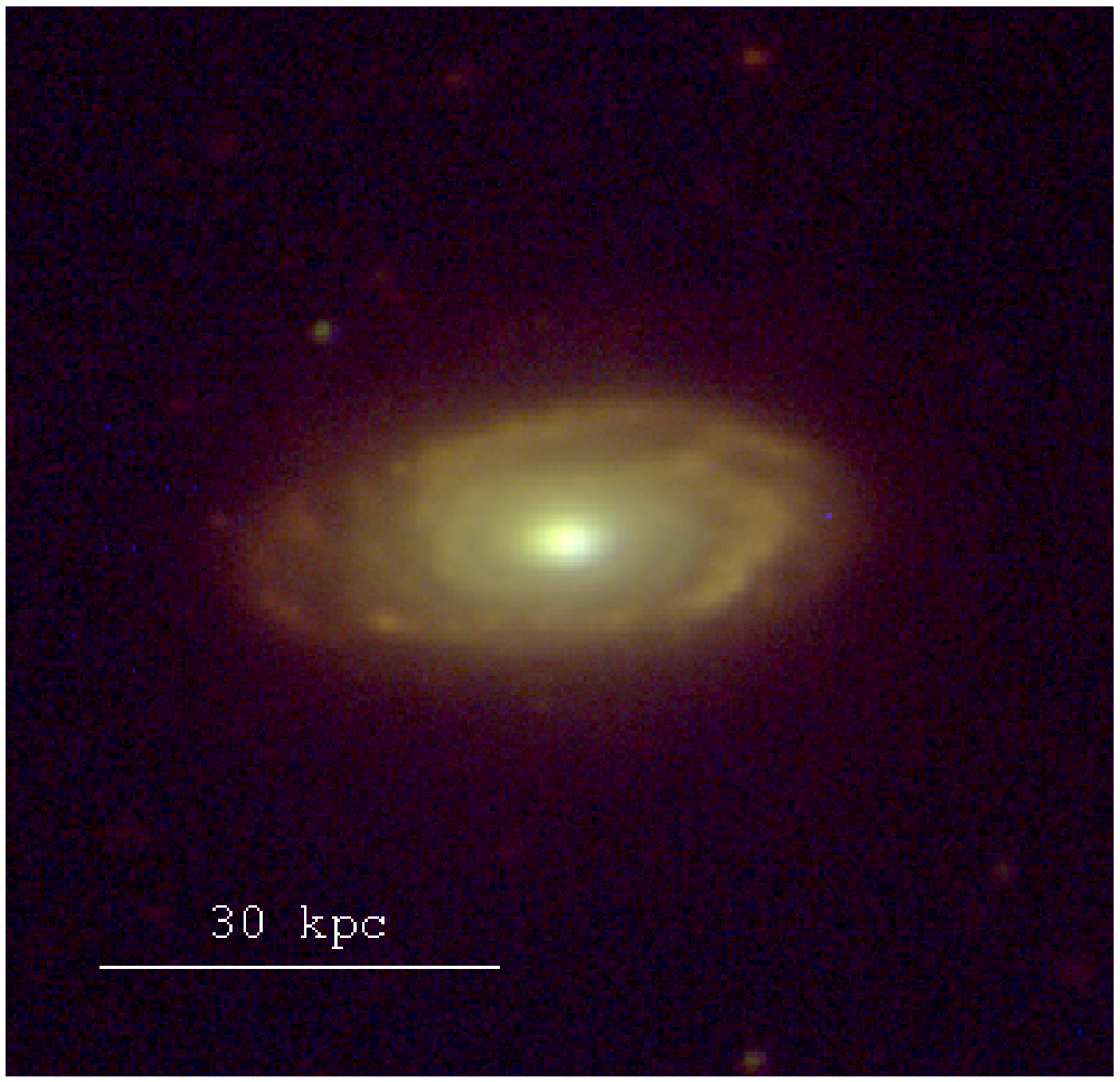}
\includegraphics[scale=0.5, angle=0]{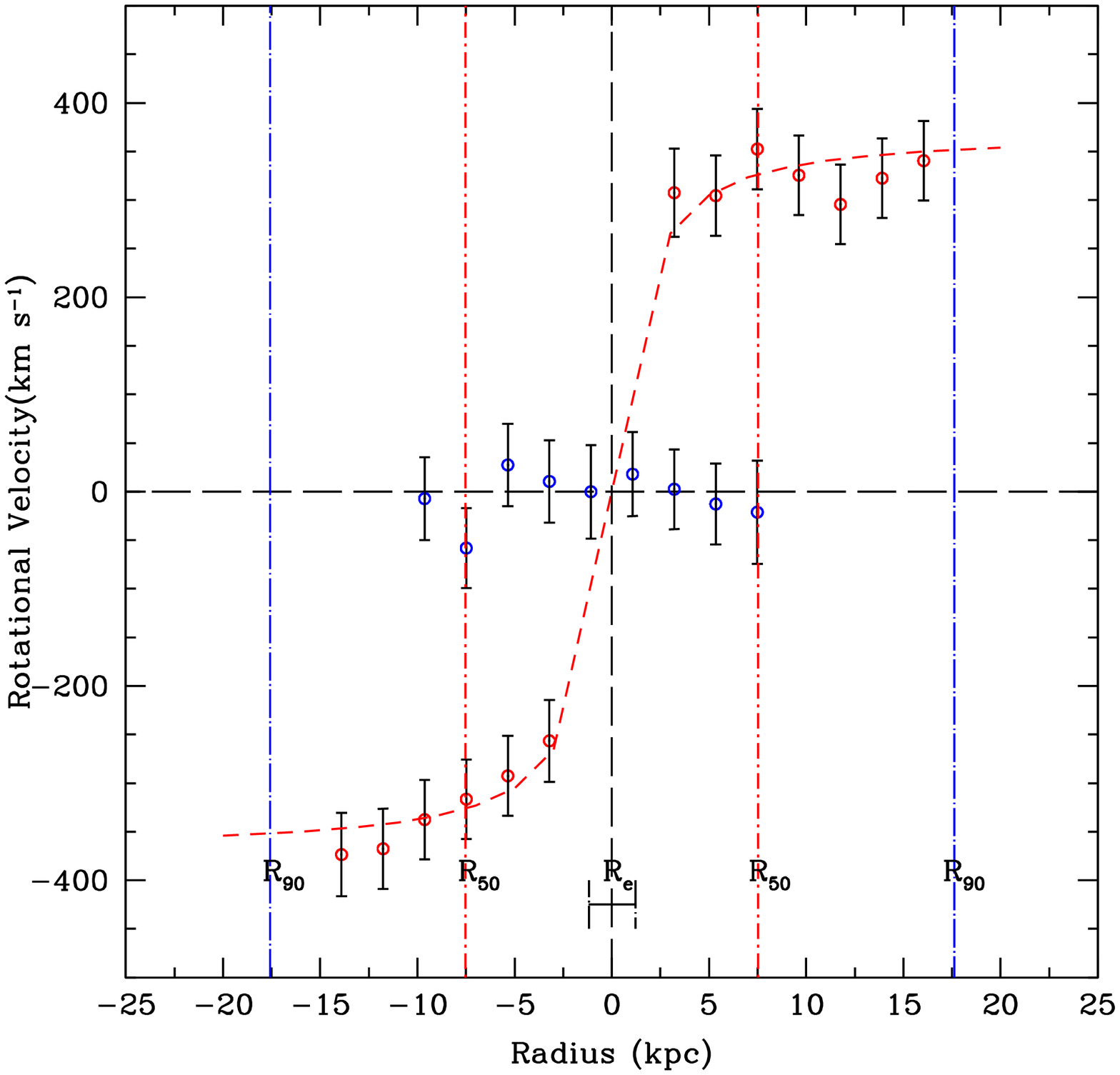}
\end{center}
\caption[]{ {\bf The optical image and the rotation curve of spiral galaxy J2345-0449.}
\footnotesize (top) A color composite  from  combining the  deep optical images
taken with the {\em MegaCam} on Canada-France-Hawaii Telescope (CFHT) in sub-arcsecond seeing.
(bottom) shows the  rotation curve of galaxy
(red dashed curve) derived from the kinematics of Balmer H${\rm \alpha}$ line
in the long-slit spectrum taken along the major-axis (red data points)
and minor-axis (blue data points). The rotational velocity is shown without the sin(i) correction for
inclination. The major axis data shows a clear
rotation signature  while  no such rotation is evident across the minor axis, clear evidence that
the emission originates in a tilted rotating disk.
For reference the  scale length of  stellar bulge component ($R_{\rm e}$)  and  the
radii containing  50\% ($R_{\rm 50}$) and 90\% ($R_{\rm 90}$) of total optical light in the $r$-band
is shown by red and blue dot-dashed vertical lines.}
\label{fig5}
\end{figure*}

We carried out the curvature correction (as above) for each of the four exposures
 and then added them after proper shifting to improve the signal-to-noise
  ratio.   We made sure that the addition is proper by looking at the
   residuals in the difference of the  frames.  We extracted
     1D spectra from sub-apertures of  5~pixel width (spatial direction) placed
     symmetrically around the  center of the galaxy. This has resulted in 15
individual spectra. The width of sub-aperture corresponds to an
      angular scale of $\sim 1.5$ arc sec (2.1 kpc), commensurate with typical
       seeing during the observing nights. In  Figure~\ref{fig4}
	    we show 2D spectrum image around the H$\alpha$
	     and [N~{\sc II}]  emission lines with the above sub-apertures marked.  
	     In the spectrum both the H$\alpha$ and [N~{\sc II}] lines clearly show the signatures 
	     of a rapidly rising and then flat rotation curve.

We simultaneously fitted the  H${\alpha}$ $\lambda 6562.81 \, \AA$ and [N~{\sc II}] $\lambda 6585.27 \, \AA$
lines using Gaussians and  $\chi^{2}$ minimization. The  second [N~{\sc II}] $\lambda \, 6549.86 \, \AA$ line 
to the blue side of H${\alpha}$ line was faint,  well within the noise, and thus we did not  
attempt to fit it. The rotational velocity $V$ at each sub-aperture position
was computed using the equation,
\begin{equation}
\centering
V = \frac{z_{\rm ap} - z_{\rm cen}}{(1+z_{\rm cen})}\,\, 3\times10^{5} \,\, \, {\rm km\,s^{-1}} 
\end{equation}
and plotted  against the distance from nucleus to  obtain the
rotation curve. Here z$_{ap}$ is the fitted redshift of  the H$\alpha$ line in a
given aperture,  and  z$_{cen} = 0.0755$, the redshift at the dynamical center of the galaxy. 
The 2D extraction is carried out for the minor axis  in the same manner as for the major axis data.

In  Figure~\ref{fig5}, we show the derived rotation curves for
the major (red data points) and minor (blue data points) axis data.  The major axis data shows a clear
rotation signature  while  no disk rotation is evident along the minor axis. This is a
 clear evidence that the emission lines originate in a tilted, rotating disk. The rotational curve shown 
 in Figure~\ref{fig4}  is fitted with an analytical function \cite{Courteau97} 
 \begin{equation}
  \centering
   V(R) = V_0+\frac{2}{\pi} V_{\rm c,obs} \times {\rm arctan}^{-1}\left(\frac{R}{R_{\rm turn}} \right)
    \end{equation}
     where, $R$ is radial distance, $V_0$ is the systemic velocity, $V_{\rm c,obs}$ is the asymptotic
      circular velocity and $R_{\rm turn}$ is the turnover radius. The $\chi^2$ minimised 
      fitting led to the  following values:
        V$_0$ = 0 km~s$^{-1}$, V$_{\rm c,obs}$ = 370.8$\pm$25.8 km s$^{-1}$
	 and R$_{\rm turn}$=1.4$\pm$0.7 kpc. Finally, The inclination 
	 corrected  circular velocity 
	 is derived at V$_{\rm c,incl}$ = 429.3$\pm$25.8 km s$^{-1}$.  We have derived the inclination
angle ($i\,$$ = 59.7^\circ \pm 2^\circ$) using Hubble's standard formula
\begin{equation}
  \centering
  sin(i) = {\left[\frac{(1 - {q^{2}})}{(1 - q_{0}^{2})}\right]}^{1/2}
   \end{equation}
where $q = (a/b)$ is the apparent axial ratio of the galaxy, as obtained 
from the fitted {\em r}-band semi-minor and semi-major  isophotal diameters of the disk 
(Table~\ref{tab:sersic_param}), and $q_{0}$ is
the intrinsic axial ratio, i.e. the ratio of vertical to horizontal scale-heights of the disk. We have adopted
a standard value $q_{0} = 0.19$, although it varies slightly from disk to disk.
	 
\subsection{\bf The Central Velocity Dispersion\,\,}
\label{veldisp}
The velocity dispersion measurements were performed using the {\em pPXF} (penalised PiXel Fitting) 
method \cite{Cappellari_Emsellem}. The {\em pPXF} method extracts the line of sight velocity distributions  
of the stars from the observed galaxy spectra, parametrized using a Gauss-Hermite series.
The code uses a penalty function that is derived from the integrated square deviation of 
the line profile from the best-fitting Gaussian. The fit is iterated until the  penalty 
function cannot decrease the variance of the fit any further. The algorithm finds the mean velocity 
$V$ and velocity dispersion $\sigma$ which minimizes the difference between the 
observed galaxy spectrum and the spectrum of a stellar template convolved by the 
corresponding Gaussian. We used the single-age, single-metallicity, stellar population library  
of  \cite{Vazdekis99} as the template.  Best-fit values of $\sigma$ from fitting the   Mg$_{b}$ 
($\lambda$ 5175 \AA) line, after correcting for the redshift and instrumental broadening, 
are $\sigma_{*} = 379 (\pm 25)~{\rm km\, s^{-1}}$
with the slit oriented along the major axis, and $\sigma_{*} = 351 (\pm 25)~{\rm km \,s^{-1}}$  along 
the minor axis. Both values are spatially averaged over  5 pixels  (2.35~kpc)  centered 
on the galactic nucleus. 

\subsection{\bf Central AGN, Star Formation Rate and Metallicity}

To decouple the possible AGN lines from the galaxy spectrum, 
we extracted the nuclear spectrum of only the central 5 pixels.
In the nuclear spectra, a narrow [N~{\sc II}] line ($\lambda$ 6583 \AA\,) of FWHM $\sim 400$ km/s is 
detected,   while the H${ \alpha}$ $\lambda$ 6563 \AA\, line is very weak and not reliably detected. 
Apart from  these features we find no other definitive signature of AGN in the optical spectrum.
The luminosities  are:  $3.02 \pm 0.27 \times 10 ^{40}$~erg s$^{-1}$  for the [N~{\sc II}] and
$< 1.05 \times 10^{40}$~erg s$^{-1}$ (3$\sigma$) for H${ \alpha}$ line. The [N~{\sc II}]/H${\alpha}$ 
line ratio lies  above that for   HII regions in the BPT diagnostic diagram
\cite{BPT} and it is  a distinct signature of ongoing AGN activity. The  small  H${\alpha}$ luminosity 
shows that this AGN is a low-luminosity active galactic nucleus (LLAGN/LINER type), which 
are defined by having an H${ \alpha}$ luminosity smaller than $ 10 ^{40}$~erg s$^{-1}$ and weak
emission lines, further indicating a very small mass accretion rate  compared
to the Eddington rate (equation~\ref{eq:eddington}).

\begin{figure*}
\begin{center}
\includegraphics[scale=0.5,angle=-90]{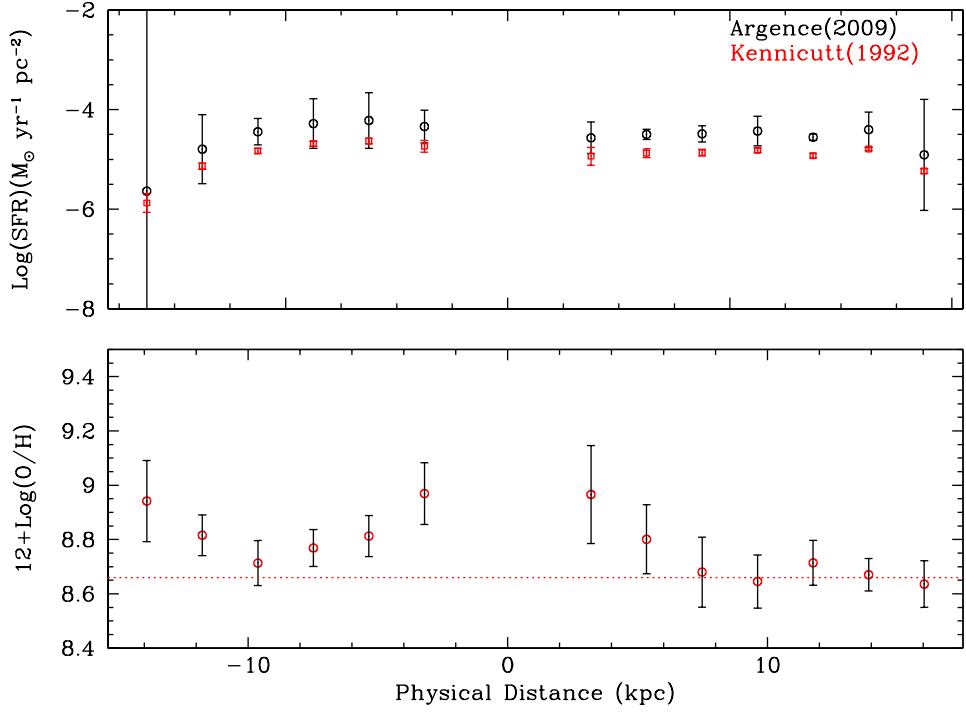}
\vspace{0.6cm}
\caption[]{\small 
The surface star formation rate (SFR) computed at different radial positions along the
major axis of the galaxy is shown in the top panel. The circles and squares are
the values computed using the prescription given by Argence \&  Lamareille \cite{AL} 
and Kennicutt \cite{Kennicut} respectively. Here we have not applied any correction for dust
extinction. The derived Oxygen  abundances  are shown in the
lower panel. The red horizontal dashed line marks the solar Oxygen abundance. }
\label{fig6}
\end{center}
\end{figure*}

For the disk extended emission, from the fitting of 2D spectra, we computed 
the integrated flux in the H${\alpha}$ and [N~{\sc II}] lines 
at different locations of the galaxy along the major and minor axes. The  H${\alpha}$ fluxes were used to
determine the star formation rates (SFR) and the H${\alpha}$ to [N~{\sc II}] flux ratios were used for calculating
the metallicities. 
For estimating the SFR
we used the equation
\begin{equation}
\centering
log({\rm SFR}) = \epsilon_{H{\alpha}}log L(H{\alpha}) - log(\eta_{H{\alpha}})
\end{equation}
where, $\epsilon_{H{\alpha}}$ and $\eta_{H{\alpha}}$ values were taken from 
\cite{AL} where  dust correction is assumed to be negligible. The surface star formation rates are
obtained by dividing the calculated SFR values by the area over which the extraction is done.
For metallicity, we have used the formula \cite{Petini&Pagel},
\begin{equation}
\centering
12+log(\rm O/H) = 8.9+0.57\times log([N{\rm II}]/H{\alpha})
\end{equation}
where O/H is the ratio of Oxygen to Hydrogen abundance,  a proxy for metallicity.
The variation of SFR and metallicity along the radial direction from centre is shown in  Figure~\ref{fig6}. 
While the surface SFR  remains virtually constant,
a radial gradient in metallicity across the disk is clearly evident. 
This indicates \sp is a  spiral galaxy  having a  large  SFR. Also notable is a striking 
 lack of recent star formation activity near the galactic center, that is possibly attributed
to  energetic feedback from the central AGN, as   discussed below.

\subsection{\bf {\em WISE} mid-infrared (MIR) colors\,\,}

The spiral galaxy \sp is  strongly detected  
in the mid-IR bands centered at 3.4, 4.6, 12 and 22 $\mu$m (W1, W2, W3 and W4 bands), observed
by the {\em Wide-field Infrared Survey Explorer (WISE)} satellite \cite{Wright10}. The  S/N ranges
from 47.9 in 3.4 $\mu$m band to 10.9 in 22 $\mu$m band. The MIR magnitudes, luminosities and 
their errors  are listed in Table~\ref{tab:infrared}. The very high MIR luminosity of the 
galaxy is unusual and suggests that it is actively 
forming stars and still building a massive stellar disk. We use the MIR colors to obtain the 
properties of a possibly dust obscured galactic
nucleus and determine the radiative efficiency of the AGN. MIR  data are highly suitable for this purpose
because the optical-UV radiation from the AGN accretion disk is absorbed by the putative dusty-torus and re-radiated
in MIR bands. It has been shown that WISE MIR colors can effectively  distinguish AGN from star forming and 
passive galaxies and, moreover, within the AGN subset itself high-excitation radio galaxies (HERG) 
and low-excitation 
radio galaxies (LERG)  stand out on the MIR color-color and MIR-radio plots \cite{Stern12,GHJ11}. The  
MIR colors for \sp are:

\begin{equation}
\centering
[W1 - W2] = 0.064, \, [W2 - W3] = 2.817, \, [W2 - W4] = 4.722 
\end{equation}

From the blue [W1 - W2]$<< 0.6$ 
and red [W2 - W3]$\sim$3 colors we conclude that \sp is a star-forming spiral galaxy 
and  there is negligible re-radiated MIR flux (no MIR `bump') 
from the accretion disk \cite{Wright10,Stern12}. This is further corroborated by the 
position of \sp on the [W3]-$L_{151}$ and [W4]-$L_{151}$  planes.  Here $L_{151}$ is  151~MHz radio 
luminosity, which is  estimated  from  the  radio spectrum and using the 
spectral index value $\alpha=2$ as calculated in Sec.~\ref{radio_prop}. On these planes 
 HERGs can be clearly distinguished from LERGs \cite{GHJ11} and \sp is  located 
in the region of LERGs (plot not shown here).  
Clearly,  \sp is a low-excitation radio galaxy  hosted by a spiral galaxy, presently its black hole is 
possibly  residing in a low radiative efficiency 
(ADAF)  state,  as also  indicated by the weak    H${\rm \alpha}$   and 
the lack of  broad emission lines in the  optical nuclear spectrum (Figure~\ref{ifosc_spect}).

\subsection{\bf CFHT Imaging and Photometric Analysis: Bulge-Disk Decomposition\,\,} 
\label{CFHT}

Modelling the light distribution in terms of  bulge and the disk components of the spiral galaxy \sp  
is important for understanding the formation and evolution of 
this highly unusual galaxy and its central black hole. 
For this purpose we used the excellent quality  archival data taken with the {\it MegaCam} wide-field imaging camera 
on the 3.6 meter Canada-France-Hawaii Telescope \cite{bou03}. These images were processed,
stacked and calibrated using the {\it MegaPipe} pipeline \cite{gwy08} at the Canadian
Astronomical Data Centre.  We used  the  {\it g} and {\it r} band images taken in excellent seeing
conditions of 0.75 and 0.80 arcsec (FWMH)  and having 
 effective exposure times of 960 and 1320 seconds respectively.  The {\it g}-band image was generated by 
 stacking eight images taken during 2008 September 01 to 2009 December 13. Eleven images taken 
 during the same period  are used to obtain stacked {\it r}-band image. The individual images were 
 median stacked using {\it SWarp} program \cite{Bertin02}.  The pixel
scale of the stacked images  was 0.18 arcsec.  The {\it MegaPipe} also provided the
inverse variance weight map for the stacked images. For additional tests and
consistency check we also used the {\em g, r, i} and {\em z} band images from the SDSS.

For obtaining the bulge and disk parameters we used {\em PYMORPH} toolbox \cite{Vinu10} 
which has  a Python based pipeline for two dimensional bulge-disk decomposition. 
{\em PyMorph} uses SEXTRACTOR \cite{ber96} and   GALFIT \cite{pen02} for object detection and decomposition. 
Briefly, we created a cutout of the galaxy based on the
    half light radius of the object. The cutout size is selected wide enough to
      include enough sky pixels to model the sky background accurately. This size is set to
      ten times the half light radius for the current case. We found that the fitting 
      results do not change significantly with the cutout size. After making the cutout image we
        generated a mask image which  masks all the bad pixels and neighbouring
	 objects during the fitting. The galaxy under study does not have any close
	  neighbours (Figure~\ref{fig3}), and therefore the contamination from neighbours 
	   is minimal. Next,  we created a configuration file for GALFIT. The configuration file includes
	    optimal initial values for the parameters of both bulge and disk. We derived the
	     initial values using SEXTRACTOR parameters. We set the total bulge and
	      disk magnitude to the MAG\_AUTO parameter and the radius to the half light
	       radius of the galaxy. The ellipticity and position angle are derived from the
	        ELONGATION and THETA\_IMAGE. A  detailed description of the procedure can be
		 found in \cite{Vinu10}.

We used a composite of the S\'ersic and an exponential function to model the galaxy light profile, where the 
former models the bulge  and the latter models the disk. The form of the S\'ersic function is

\begin{equation}
\Sigma(r) = \Sigma_e \exp\left(-\kappa\left[(r/r_e)^{1/n} - 1\right]\right)
\end{equation}
where $\kappa$ is a function of S\'ersic index ($n$). For $n = 4$, $\kappa = 7.67$
which corresponds to the de Vaucouleurs law. $\Sigma_e$ is the surface
brightness at $r_e$, the scale radius (half light radius). The disk component is
modeled using the exponential law which is given by
\begin{equation}
\Sigma(r) = \Sigma_0 e^{-\left(r/r_d\right)}
\end{equation}
where $\Sigma_0$ is the central surface brightness and $r_d$ is the disk scale length.
The half light radius ($r_{e}$) is related to disk scale length as                                                     $r_{e} = 1.68 \times r_d$ and $\Sigma_e = \Sigma_0 / 5.36$.

\begin{figure*}
\begin{center}
\includegraphics[scale=0.5,angle=0]{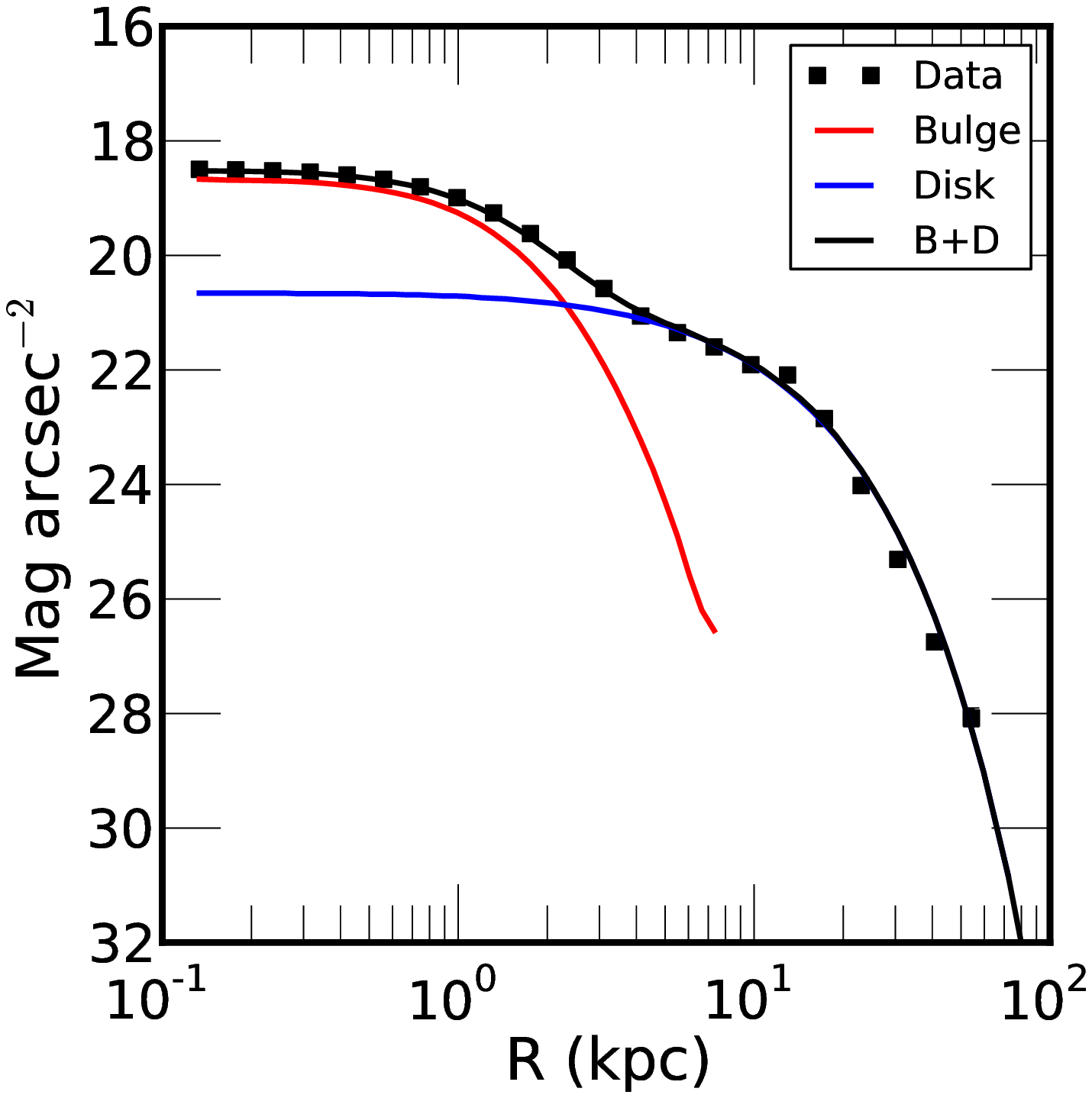}
\includegraphics[scale=0.5,angle=0]{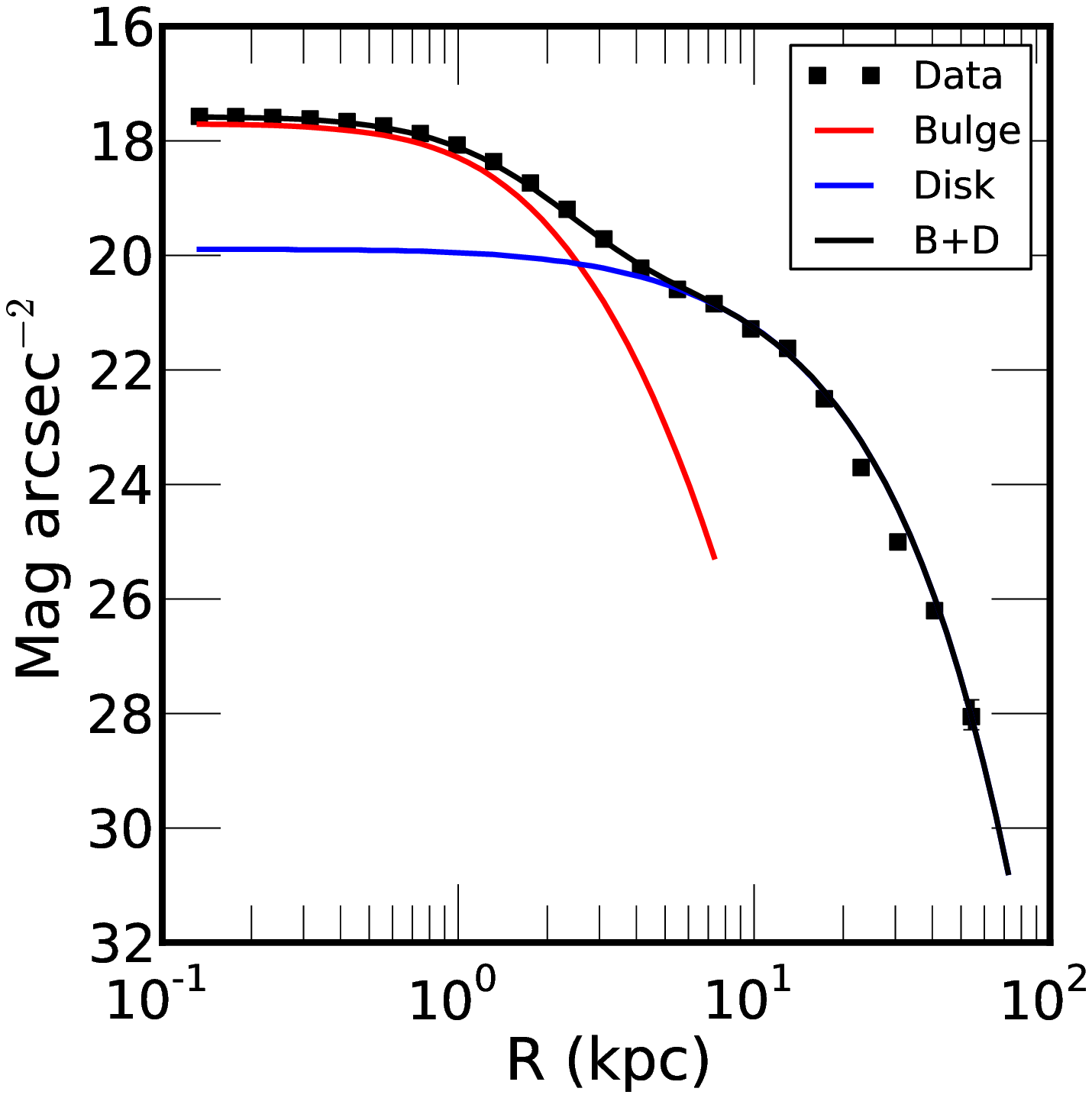}
\vspace{0.6cm}
\caption[]{\small One dimensional surface brightness profile comparison. The black squares
are observational data points obtained from CFHT image in {\it g}-band (left panel)
and in {\it r}-band (right-panel).
Red and blue lines are the model profiles for the bulge and disk components respectively.
The black line shows the combined profile  by adding the bulge and disk parts. The model profiles were obtained 
from a point spread function (PSF)  convolved image generated by  the GALFIT software, as described in the
main text.}
\label{fig7}
\end{center}
\end{figure*}

The bulge-disk decomposition was done by $\chi^2$ minimization. In this technique
the observed image is compared to a PSF convolved linear combination of
S\'ersic and exponential functions. Therefore, it was necessary to model the
point spread function (PSF) well in order to reliably estimate the photometric parameters. We
used   stellar images closer to the galaxy to represent the PSF. For that we
selected stars from the CFHT images based on the CLASS\_STAR parameter from
SEXTRACTOR and create stamp size images. These stellar images were checked
manually using the IMEXAMINE program in IRAF. This  helped to remove saturated
stars and some other stars which are contaminated by bright neighbours.
In Table~\ref{tab:result_mag} and Table~\ref{tab:sersic_param} we list the fitted values 
of  the bulge and disk parameters estimated
using the {\it g} and {\it r} band CFHT images and   the SDSS images. We found  
CFHT  results fairly consistent with
those obtained using the SDSS images, although the former were obtained under better seeing conditions. 
Although we used the pipeline, we vary the initial conditions of different parameters 
to check whether the final results correspond to a global minimum. We found that the results are stable against
different initial conditions. Finally, we checked for any possible
contribution from a possible point source at the center of the galaxy. 
We found that any such point source  must be too faint and its inclusion doesn't improved the fit.
The magnitudes shown in  Table~\ref{tab:result_mag} and Table~\ref{tab:sersic_param}  have been
corrected for foreground extinction using Schlegel et al. \cite{sch98}. We also applied
k-correction using E/S0 SED from Poggianti et al. \cite{pog97}.  Figure~\ref{fig7} shows a comparison of
 the  modelled profiles of bulge and disk components with the observed profile of the
galaxy in the CFHT {\it g} and {\it r} bands. These are generated using IRAF/ELLIPSE task from
the PSF convolved output images generated by GALFIT. It can be seen that the 
bulge-disk model agrees very well with the observed  profile of the galaxy. 

The S\'ersic index for the bulge is found to be
 small at $n \approx 1$, which essentially implies  a  `pseudo-bulge' with a disk-like  exponential 
light profile ($n = 1$ gives a pure exponential profile,
and  $n = 4$ corresponds to the  de Vaucouleurs profile).  Moreover, the
scale length of the  bulge is  very small compared to the  disk ($r_d \approx 7 \times r_e$) 
and  only about $14 - 18\%$ fraction of the
total {\it g}-band or  {\it r}-band luminosity is contributed by the bulge.  This again indicates
that the \sp spiral host contains  a pseudo-bulge   rather than  a classical bulge component.  
Galaxies with  classical bulges generally  have a  much more centrally peaked
light profile,  contain a higher fraction of total light and their S\'ersic index  is
larger ($n \approx  2 - 6$), as compared to pseudo-bulges which have $n <  2$ 
\cite{Fisher_Drory,Kormendy_etal}.

\subsection{\bf Additional checks to confirm the pseudo-bulge nature of \sp\,\,}
\label{checks}

\begin{figure*}
\begin{center}
\includegraphics[scale=0.6,angle=0]{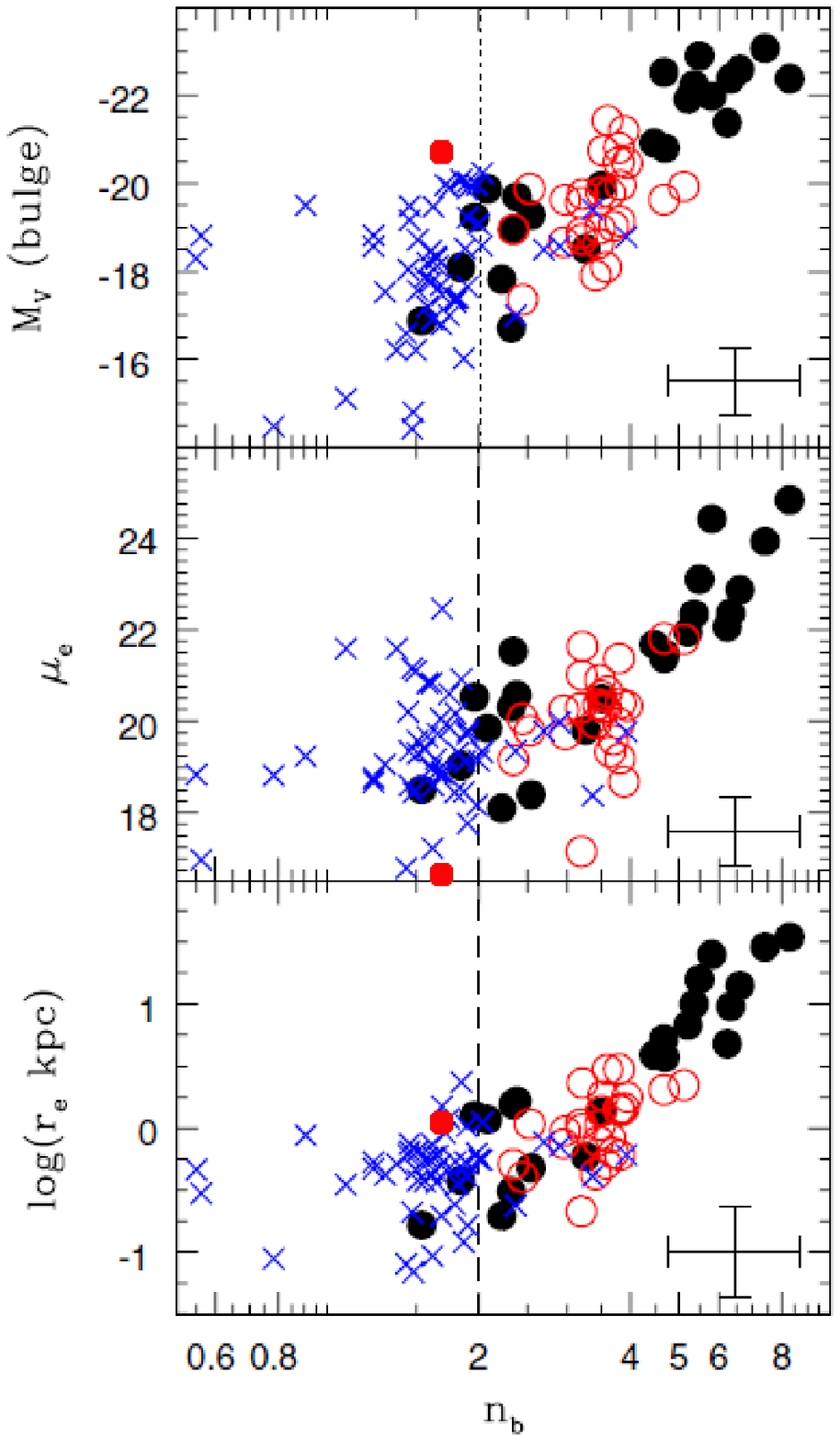}
\includegraphics[scale=0.7,angle=0]{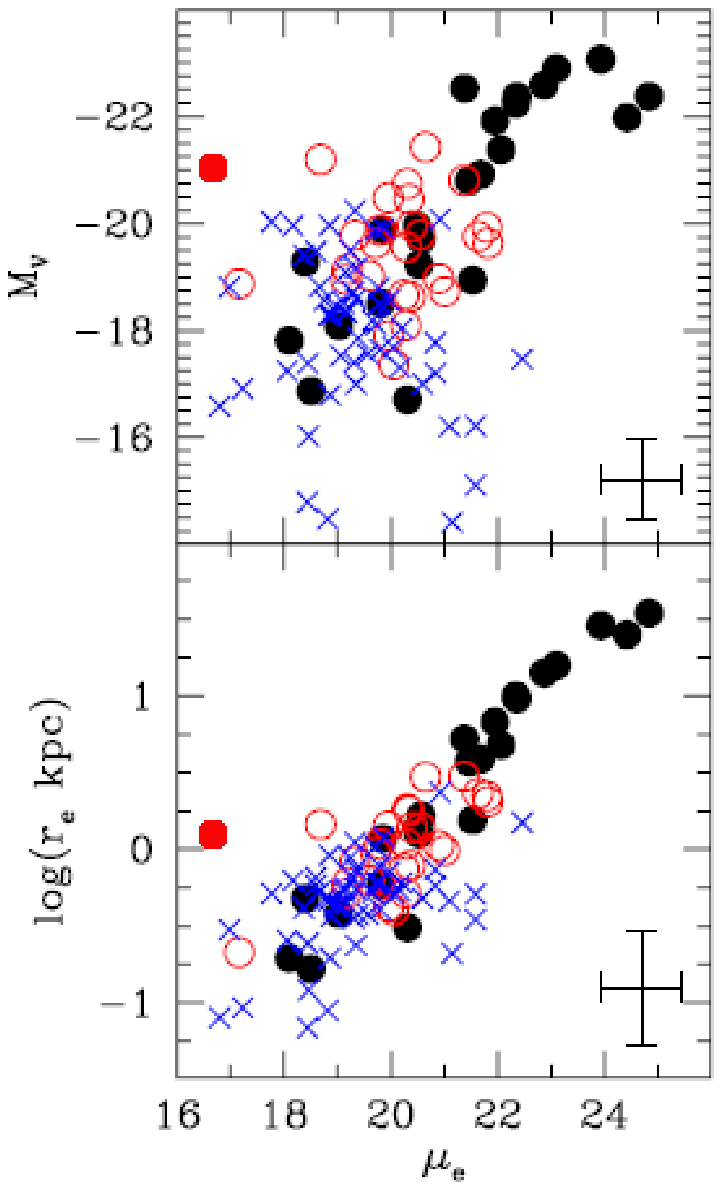}
\includegraphics[scale=0.7,angle=0]{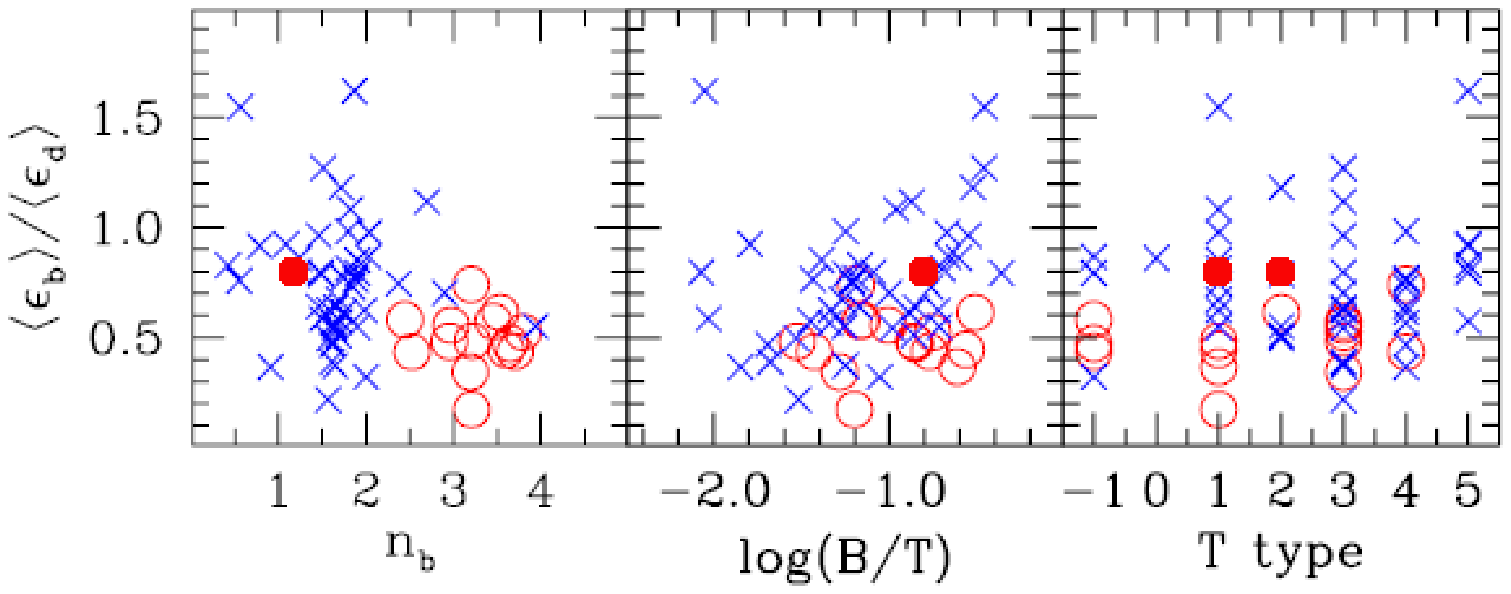}
\vspace{0.6cm}
\caption[]{\small These figure panels  show the comparison of structural and photometric
properties of the bulge in  present spiral galaxy \sp with  bulges in other galaxies (both classical and pseudo bulges)
as shown in  \cite{Fisher_Drory}. Here $n_{b}$, $M_V$, $r_e$ and $\mu_e$ denote the S\'ersic index,
absolute V magnitude, scale radius and average surface brightness within $r_e$ of bulge respectively.
The dashed line in the first figure panel separates the classical bulges ($n_{b} > 2$) from 
pseudo bulges ($n_{b} < 2$).
$\epsilon_b$ and $\epsilon_d$ are the ellipticity of bulge and disk components, and B/T is bulge to total
light ratio. T-type is the morphological type of the galaxy (see main text). 
Blue crosses and red open circles represent the pseudo-bulges and classical bulges respectively in spirals,  
the filled red dot represents the present galaxy J2345-0449, while black filled points correspond to the ellipticals.
\sp clearly belongs to the pseudo bulge category as seen on  most of these  plots. {\bf We show two red dots in 
last figure panel; one for T-type = 1 (S{a}) and other for T-type = 2 (S{ab}). This is to take care of 
the uncertainty in the visual T-type of the present galaxy}.
}
\label{fig8}
\end{center}
\end{figure*}

We have compared several photometric parameters of the (pseudo) bulge in \sp  to some
other well known previous studies  in order to  further verify its  pseudo bulge 
classification. According to  \cite{Gadotti09},
pseudo bulges have small S\'ersic index and scale radius and they occur in
galaxies having low bulge to total light ratio (B/T). Our estimated values of these parameters 
(Table~\ref{tab:result_mag} and \ref{tab:sersic_param}) fall
very close to those of pseudo bulge systems in  \cite{Gadotti09}.  We further 
compared our values with those of galaxies studied by   \cite{Fisher_Drory}. 
In  Figure~\ref{fig8} we show that our fitted values for \sp
fall near the pseudo bulge galaxies in $n_{b}-M_V$, $n_{b}-\mu_e$ and $n_{b}-r_e$ planes of
  \cite{Fisher_Drory}. Similar
trends were found in the $\mu_e - M_V$ and $\mu_e-r_e$ planes. Here $n_{b}$, $M_V$, $r_e$ 
and $\mu_e$ are the S\'ersic index,
absolute {\em V} magnitude, scale radius and the average surface brightness within $r_e$, respectively.
It is clear that \sp is  located closer to pseudo bulges and far from classical bulge systems.
However,  the  brightness  of the  bulge and $r_e$ in \sp is higher  than that shown for 
pseudo bulge disk galaxies in  Fisher and Drory \cite{Fisher_Drory}.  
Further evidence comes from  the  location of  \sp in
the $\langle \epsilon_b \rangle / \langle \epsilon_d \rangle - $ B/T plane (Figure~\ref{fig8}),
where $\epsilon_b$ and $\epsilon_d$ are the
ellipticities of the bulge and disk. We found that our values for \sp closely matches with
those of the pseudo bulge population, and 
 the ratio of ellipticities ($\langle \epsilon_b \rangle / \langle \epsilon_d \rangle$)
is higher than for classical bulges in galaxies of
similar morphological T-type\footnote[4]{T-type is a numerical value assigned  visually to each class of galaxy in  
the De Vaucouleurs morphological classification scheme. T values run from -6 to +10, 
with negative numbers corresponding to early-type galaxies 
(ellipticals and lenticulars) and positive numbers to late types (spirals and irregulars). Spiral galaxies 
are assigned to a class based primarily on the tightness of their spiral arms, with larger  T-value representing
more open arms.}. 
All the above checks  strongly suggest that the central bulge in \sp
should  be classified as a  pseudo-bulge, rather than a classical bulge.
Consequently,  evolution of the \sp host galaxy is  probably governed by  by  quiet,
`secular evolution'  processes  in which disk instabilities drive gas to
the center, rather than violent  merger events  that would have led to a classical bulge \cite{Kormendy_etal}. 
This inference is  further bolstered  by the location of \sp in a sparse galactic environment,
absence of  any visible tidal debris (like tails, plumes or shells)  and its well-developed
spiral arms within a stable, rotationally  supported stellar disk.


\section{RESULTS: A SPIRAL HOST WITH EXTRAORDINARY PROPERTIES}
\label{results}
The extreme  nature of this spiral galaxy  is  underscored, firstly, by its
unusual rotation curve. Rotation curves are the major tool for determining the mass distribution 
in spiral galaxies, e.g. \cite{Sofue_Rubin}. The kinematics of the  Balmer H${\alpha}$  line
in the long-slit spectrum along the major axis, taken with the IFOSC on 
{\em IUCAA Girawali Observatory}  2m telescope, 
shows a rapidly rising rotational velocity in the inner $\sim$1 kpc,  attaining an extremely large
 value,  $V_{\rm rot} \sim  430\,~{\rm km s^{-1}}$ (inclination corrected)
in the asymptotic flat region at $r \gtrsim 10$~kpc from the galactic center (Sect.~\ref{rotcurv}
and Figure~\ref{fig5}).
No kinematic signature of  rotation is detected along the minor axis, which is a clear evidence for a planar,
rotating galactic disk. Such rapid disk rotation ($V_{\rm rot} \sim 400 - 500\ ~{\rm km s^{-1}}$) 
is rare and has been reported earlier for  just a few
exceptionally  massive spirals \citep[e.g.,][]{Giovannelli_86,Rigopoulou_2002,Akos13}. Assuming a
quasi-spherical halo,  the  dynamical mass (luminous plus dark-matter) enclosed
within a radius $r$ is $M_{\rm dyn} = \kappa r {V_{\rm rot}^{2}}/G  $, where
$\kappa$ is a  correction factor, in range $0.6 - 1$ for a
thin-disk to a spherical halo mass distribution model \cite{Lequeux},
and G is the gravitational constant.  We estimate 
$M_{\rm dyn} = 1.07 (\pm 0.13) \times 10^{12} M_{\odot}$,
for $\kappa = 1$ and  $V_{\rm rot} \sim 430~{\rm km s^{-1}}$  at $r = 25$~kpc, i.e.
$\sim$3.5 times the scale radius of the exponential disk. This places \sp amongst the brightest and
most massive spirals known, having  $\sim7$ times the mass of the Milky Way
within an equivalent radius, using the Tully-Fisher relation. Furthermore, we obtain its 
virial mass $M_{\rm 200} \sim 1.05 \times 10^{13} M_{\odot}$  and virial radius 
$R_{\rm 200} \sim 450\,~{\rm kpc}$ using 
the baryonic Tully-Fisher relation (BTFR) for  $\Lambda$CDM cosmology \cite{McGaugh05} 
 ($M_{\rm 200} \propto V_{\rm rot}^{3.23}$) and scaling
from $M_{\rm 200}$ and $V_{\rm rot}$ ($ 448~{\rm km s^{-1}}$) of the fast 
rotating spiral galaxy NGC~1961 \cite{Akos13}. 

Secondly,  photometry and 2-D luminosity profile fitting
of  the CFHT and SDSS images showed that \sp is an extremely luminous (optical {\em r}-band: $M_{r} = -23.26$, near IR
{\em K}-band: $M_{K} = -26.15$)   isolated spiral galaxy (see above) 
lacking  a  spheroidal classical bulge, but having  a compact  disk-like  `pseudo-bulge' of
low S\'ersic  index ($n \approx 1$) which contributes only  $\sim14\%$ to the
total {\it g}-band luminosity. In contrast, classical bulges generally 
have a  more centrally peaked brightness  profile characterized by a larger 
S\'ersic  index $n \approx $ 2 -5 \cite{Fisher_Drory}. 
Recent recognition of a high fraction ($\sim 60 \%$) of nearby spiral  galaxies
without a kinematically `hot'
central bulge challenges the conventional ideas of  hierarchical galaxy formation;
how  so many bulgeless pure disk galaxies have come to exist, despite the galaxy
mergers known to occur \cite{Fisher_Drory,Kormendy_etal}? Below, we discuss this further, in the context
of mass estimation of the central SMBH. \sp  shows  ongoing star forming activity across the
galactic disk, traced by  H${ \alpha}$ and [N~{\sc II}] 
emission lines. It is also strongly detected by the {\em  Galaxy Evolution Explorer} (GALEX)
in the near (2316 \AA) and far UV (1539 \AA) bands \cite{martin05}, indicating
a  starburst of  massive, hot O,B stars during the past $\sim100$ Myr.
Interestingly, UV or H${\alpha}$ signatures of  recent star formation ($\sim100$ Myr) are
absent closer to the nucleus, showing only  stellar population older than $\sim1$ Gyr (Figure~\ref{fig6}). 
We interpret this as a result of  heating and expulsion of the star forming medium by
energetic, episodic feedback of the  central AGN, as revealed by  two pairs of radio lobes.


Another unusual property displayed by this disk galaxy is its exceptionally 
large stellar velocity  dispersion
of $\sigma_{*} = 379 (\pm 25)~{\rm km\, s^{-1}}$ measured along its major axis, 
and $\sigma_{*} = 351 (\pm 25)~{\rm km \,s^{-1}}$ along the minor axis, averaged within the central 2.35~kpc 
region (Sect.~\ref{veldisp}). The slightly higher $\sigma_{*}$  for the major axis possibly  
indicates some contribution from the disk rotation. Hence, we adopt  the smaller $\sigma_{*}$  
of the minor axis as representative of the galaxy's central velocity dispersion. It is noteworthy 
that this dispersion value is significantly larger  than that known for the majority of bulge-less disk galaxies
on such a spatial scale \citep[e.g.,][]{Hu,Graham,Kormendy_nature,shankar2012}, and compares well with
that usually found  for massive elliptical galaxies \citep[e.g.,][]{Haring_Rix,Marconi_Hunt,Gultekin}.  
This strongly hints at a large central concentration of mass,  including a  putative  SMBH. We 
obtain the dynamical mass of the (pseudo) bulge $M_{\rm b} = 1.07 (\pm 0.14) \times 10^{11} M_{\odot}$, 
within one scale radius (${\rm r_{e}} = 1.25$~kpc), and  a  
black hole mass $M_{\rm BH} = 2.54 (\pm 0.48) \times 10^{8} M_{\odot}$ using the black hole mass  
vs. bulge mass  correlation found  by Marconi and Hunt  \cite{Marconi_Hunt}. This value 
is comparable to another estimate of $M_{\rm BH} = 3.88 (\pm 0.40) \times 10^{8} M_{\odot}$,
  obtained from the $M_{\rm BH} - \sigma_{*}$ correlation of  Hu \cite{Hu} for the pseudo bulges.
  However if we use the  $M  _{\rm BH} - \sigma_{*}$ relation of Gultekin et al. \cite{Gultekin}, calibrated using 
  a sample of both ellipticals and spirals, we obtain the  black hole mass 
  $M_{\rm BH} = 1.43  \times 10^{9} M_{\odot}$. Similarly, using {\em K}-band $M  _{\rm BH} - L_{K}$ relation
  of Graham \cite{Graham07}, we obtain $M_{\rm BH} = 1.22  \times 10^{9} M_{\odot}$. Both these latter masses are
  close to each other, yet are significantly larger than the previous two mass estimates.  We  need to verify  
  these rather large mass  values via more direct and robust methods.  If confirmed it would imply 
  that \sp  hosts a  SMBH  as  massive as those found  in  giant elliptical galaxies, which is   
   not expected  for normal disk galaxies and is extremely unusual for  a pure disk galaxy 
   like \sp. 

\section{DISCUSSION}
\label{discussion}
  Despite lacking  a classical bulge, this exceptional spiral galaxy  is  likely to host  an
  unusually massive,  accreting (radio-loud) black hole of $ M_{\rm BH} > 10^{8}$~M$_{\odot}$. However
  our present data do not  provide strong constraints on $ M_{\rm BH}$, mainly because 
  \sp lacks a classical bulge,  whereas the black hole mass is 
  more easily obtained from the tight $M_{\rm BH} - \sigma_{*}$ correlation in bulge dominated systems
  \cite{Haring_Rix,Marconi_Hunt,Gultekin}. For this reason the demographics of black holes in disk galaxies is poorly
  determined.  However, there is growing evidence that  a classical bulge is not essential for  
  nurturing  a massive black hole  and some pseudo bulge systems  may also host fairly massive 
  black holes of masses $10^{3} - 10^{7} M_{\odot}$  \cite{Kormendy_nature,shankar2012}. Therefore 
   these black holes must somehow attain 
  masses substantially larger than the values implied by their bulge properties, possibly via a disk-driven,
  non-merger growth route.  It appears that \sp   is one such system, being an extreme  member of such a population. 
  Very little is known of the growth and AGN activity of super massive black holes  
  in bulge-less spiral galaxies. Therefore, it would be a  very significant result  
  if more detailed observations show  that \sp harbors a 
  super massive black hole of $> 10^{8}$~M$_{\odot}$. We note that some previous  
  works have  suggested   \citep[e.g.,][]{Hu,Graham,shankar2012},
  that pseudo bulge and barred galaxies  may host  black holes that are  about 3 to 10 times less
  massive than pure ellipticals having the same $\sigma_{*}$. It has also been proposed that  their black 
  hole masses do not correlate  well with properties of galaxy disks or their 
  pseudobulges \cite{Kormendy_nature}, and hence    
  the  growth process of  massive black holes in flattened galactic 
  disks lacking prominent bulges is  still  mysterious  and  a  subject  of  intense   
  debates and investigations \cite{shankar2012,Kormendy_nature,Simmons13}. 
  
  Our estimation of  black hole mass in \sp could be a lower limit,  constrained by the lack of sub-arcsec 
   resolution spectral data. Nevertheless, it is interesting to note that the 
   lower range of our estimated black hole mass  (2 - 4~$\times 10^{8}$~M$_{\odot}$)  
   is close to the minimum required for the production of large-scale relativistic jets 
   in radio galaxies and   quasars 
   \cite{Laor,G-K_M_W,Sikora07,CM11}. At the present stage one can not rule out an even more 
   massive  black hole of 
   $\gtrsim 10^{9}$~M$_{\odot}$ in this galaxy, given its extreme radio properties and exceptionally large mass. 
   Therefore, a more robust and direct  estimate of the black hole mass in \sp is needed, possibly via
  spatially resolved stellar kinematics or gas dynamics measurements close to
   its gravitational sphere of influence: 
   $r_{g}\sim50\,(M_{\rm BH}/{10^{9}\,M_{\odot}})(\sigma_{*}/{300\, {\rm km\, s^{-1}}})^{-2}\,~{\rm pc} \equiv 
   0.035^{''} - 0.35^{''}$ 
   for $M_{\rm BH} \sim 10^{9} - 10^{10} M_{\odot}$. At $\sim340$~Mpc distance ($1^{''}\equiv1.43$~kpc)
   such a refined measurement is  challenging  if    $M_{\rm BH} \lesssim 10^{9} M_{\odot}$, requiring
   very high ($\sim 0.01^{''}$)  spatial resolution. 
  Nevertheless, this method has recently  revealed a $1.7 \times 10^{10} M_{\odot}$ 
  extra-massive black hole in the nearby galaxy NGC~1277, 
  another fast rotating,  disk  galaxy with a pseudo-bulge \cite{van-Der-Bosch_Nature}.
  Interestingly, for NGC~1277 the stellar velocity dispersion is  $\sim 333\,~{\rm km\, s^{-1}}$ at  
  the half light radius ($2.8^{''}$) -- very close to the present case -- 
  which  rises to $> 400\,~{\rm km\, s^{-1}}$ near the black hole's  sphere of influence. Therefore, it is
  tempting to speculate that \sp might also harbor an extremely massive black hole, that would help
  to explain its unusual radio properties, further discussed below.



Because it is not obvious how this galaxy acquired such a range of  extraordinary properties, it
poses challenges to theoretical models  and holds  important clues for understanding
the co-evolution of black holes and  disk galaxies,  accretion physics,  the 
formation of large scale relativistic jets in AGN, and a possible solution to the long standing puzzle of why 
disk galaxies are nearly universally devoid of powerful double radio sources.
We  propose  that this  spiral galaxy's  huge  mass, angular momentum, and  episodic
radio jet ejection are all the results of
its  rare,  unusual evolutionary history that has led to the formation of both a rotationally supported
massive, fast-spinning galactic disk (here $({V_{\rm rot}/\sigma_{*}})^{2}\sim 1.5$), as well as  a
 mass accereting, possibly  rapidly spinning, unusually massive central SMBH. 

Very little is known about the formation and evolution of such extremely
massive spiral galaxies, as they are  found rarely  and hardly ever known to eject
powerful relativistic jets on Mpc scale like \sp. 
The cosmic evolution of the galactic disk of \sp  may  have been 
primarily driven  by  external cold gas accretion  and  internal `secular'
processes instead of  recent ($t \ltappeq$ few$\times10^{9}$~y) 
major merger events. This is suggested by its disk-like pseudo-bulge, 
absence of  close galactic neighbours and of
tidal debris (e.g., stellar streams, plumes or shells), and by its stable, 
rotationally supported massive stellar disk showing well formed spiral arms.
Furthermore, a close alignment  (within $\sim 10$ degree) of the inner and outer radio-lobe pairs  
implies that the black hole's spin axis  -- which in equilibrium orients itself orthogonal to the inner
accretion disk due to the Lense-Thirring frame dragging effect \cite{BP75}-- has remained stable (non precessing),  
at least between the two last episodes of jet triggering on timescales of $t\sim10^{8}$~y.
All this suggests that the  galactic disk and  its central black hole  might have a quiet evolution together, without
major mergers. The massive galactic disk is  clearly rotationally supported ($({V_{\rm rot}/\sigma_{*}})^{2}\sim 1.5$) 
and should be dynamically stable against gravitational instability. 

The formation epoch of \sp  probably also lies  
in the  early  universe  (redshifts $z \gtrsim 2 - 3$)
more than $10$~Gyr ago, i.e. during the era of galaxy formation and quasar activity. In that era,  
most galactic disks (pancakes) were unvirialized, still assembling around isolated  halos 
via minor mergers and mass accretion from  cool, gas-rich filaments of the cosmic-web, and
often experiencing high star formation rates of $\sim10^{2} - 10^{3}$~$M_{\odot}{\rm yr^{-1}}$.
The  initial build up of the present large large mass and angular momentum  of the galactic disk  of \sp 
possibly started  very early  (probably $z \gtrsim 6$) during the pancake stage of its parent halo,   
via gravitational tidal torques    and rapid  disk feeding  through a few
dominant, coplanar cold-streams (`cold mode' accretion),  as recently observed for a star forming disk 
galaxy at  $z = 2.3$ \cite{Bouche13} and also shown  in numerical 
simulations \cite{Keres05,Governato07,Dekel09,Dekel012}. For very massive halos of mass $\gtrsim 
10^{12} M_{\odot}$, another  `hot mode' growth also takes place via  the intergalactic
gas accreting quasi-spherically, 
experiencing an accretion shock and heating to virial temperature as it 
collides with the hot, hydrostatic halo near the virial radius, and then being accreted onto the
galaxy on radiative cooling time scale \cite{Keres05,OcVirk08}. 
Simultaneously, the growth of the massive black hole of this bulge-less spiral  
possibly started from an initial `seed'  black hole mass, via an 
internal disk-driven (secular) accretional route - through a  rapid phase of mass and angular momentum (spin)
transfer from a gas-rich accretion disk fed at high Eddington rate \cite{BV08}. Its inner accretion zone was 
coupled to the fast-spinning outer galactic disk via the viscous and magnetic torques.  
This coherent growth process would  spin-up the  growing black hole  to a high value 
if copious fuel supply is available for a long enough time; in contrast to the
low  final spin  expected in  a stochastic (i.e. chaotic or incoherent)
merger driven growth scenario \cite{Dotti13,Dubois14}. It is well known that a  black hole approximately doubling 
its mass by accreting a constant angular momentum gas from a standard thin-disk  would
acquire a maximum `canonical' spin  of  $a = 0.998$ \cite{Thorne74} due to a braking torque, where the    
dimensionless spin parameter is $a = c J_{\rm BH}/G M_{\rm BH}^{2}$ and black hole's 
angular momentum is $J_{\rm BH}$. Such a massive, acccreting-spinning black hole  would be a
highly efficient `engine' for launching, fast,  collimated plasma jets  via the 
Blandford-Znajeck mechanism (or  variants of it),  further boosted by a 
strong build up of magnetic flux  near the black hole horizon  by advection of the accretion flow. 
Below we  further comment on  the feasibility and implications of such a SMBH growth and jet formation 
process in \sp.


Obviously, the most striking property that sets \sp apart from nearly all other  spiral galaxies 
(including Seyferts) is its large mass and  unusual radio loudness, resulting from
highly efficient ejection of 
relativistic jets on Mpc scale. The extreme rarity
of such galaxies implies that whatever physical process  that created such huge radio jets in \sp must be very 
difficult to realize and maintain for long periods of time in  most other spiral/disk galaxies. 
Thus an important question is; what could be responsible for the efficient fuelling and sustained collimated jet 
ejection  activity in \sp? The answer requires a  very detailed knowledge of the 
properties of its central black hole engine, 
i.e. its mass, spin and the nature of  its accretion  flow, viz., the  mass accretion rate and
the magnetic field topology of the inner accretion disk, most of which  is clearly lacking at present. 
Although the exact mechanism of jet launching is not known for \sp, a promising and widely 
accepted explanation  is the  Blandford-Znajek   process  that efficiently
extracts  the rotational energy of the black hole and  
the orbital energy of the accretion flow into an intense outflow of  Poynting flux dominated 
electromagnetic jet. In the Blandford-Znajek  process the jet launching requires 
a massive, spinning Kerr black hole, and an accretion disk threaded 
with strong poloidal  magnetic field lines advected near the hole's outer horizon 
at radius $R_{\rm H} = (G M_{\rm BH}/{c^{2}}) [1 + (1 - {a^{2}})]^{1/2}$. 
In the classic BZ process the electromagnetic luminosity of jet $\bar{Q}$ scales with black hole properties as 

\begin{equation} 
\centering
\bar{Q}_{44} \approx  0.2\, M_{8}^{2} \, a^{2} \, B_{4}^{2}, 
\label{eq:BZ}
\end{equation}

where, $\bar{Q}_{44} = \bar{Q}/10^{44}~{\rm erg~s^{-1}}$, the
black hole mass  $M_{8} = M_{\rm BH}/10^{8} M_{\odot}$ and  the poloidal magnetic field $B_{4} = B/10^{4}$~Gauss
\cite{BZ77,MT82,TNM10,SNPZ13,PNS13}. Thus, the jet power depends strongly on the black hole mass, spin and  
the magnetic flux at the horizon. This prediction is confirmed by 
recent numerical general relativistic magnetohydrodynamic (GRMHD) simulations  
which show that the efficiency ($\eta$) of 
conversion of accretion luminosity to the jet luminosity (including a possible coronal wind)  could be as 
high as $\eta = (\bar{Q}/{\dot M} c^{2}) \sim 30 - 140\%$ for  spin $a = 0.5 - 0.99$, if the 
magnetized accretion flow manages to  accumulate  (via advection and turbulent amplification) 
large magnetic flux near the black hole's horizon 
\cite{TNM10,TNM11,PNS13}. Similar results have been found by Hawley and Krolik \cite{HK06} which  demonstrate
that the electromagnetic extraction of the spin energy of a black hole is very much feasible.

Observationally, the jet kinetic power ($\bar{Q}$) is a key descriptor of
the state of  an accreting SMBH system; its mass, spin and the magnetic field of the accretion disk.
Correlation of low-frequency ( $\nu \sim150$~MHz)  radiative radio power  of  FRII radio 
sources  against their jet power shows 
that the radio luminosity of the jet constitutes only a small fraction ($<1\%$)
of the total  kinetic power output \cite{Punsly11,Daly2012}. Here we use the low frequency radio flux, 
which usually originates in diffuse, optically thin radio lobes moving at low velocities, thus the relativistic
 beaming effects are negligible. We estimate the 151 MHz lobe flux density to be  15.75~Jy (excluding the inner
 jet/lobe pair and the core) using the 0.3 GHz and 1.4 GHz measurements 
 and  the observed radio spectral index  $\alpha=2$. Using 
 the  published calibration relations  \cite{Punsly11,Daly2012}  we obtain
 $\bar{Q} \approx  1.7 \times 10^{44}$erg~s$^{-1}$ for the (time-averaged) 
jet kinetic power in \sp. A similar value of $\bar{Q}$ is obtained  using the 
relation  for estimating the total jet power from radio power, 
 for a $M_{\rm BH} \sim  10^{8} M_{\odot}$ \citep[equation 1 in][]{meier01}.
As we have applied no correction for the (unknown)  energy loss in the outer radio lobes,  
$\bar{Q}$ is likely to be a lower limit. This high value  of $\bar{Q}$ in \sp is 
typical  of high powered FR~II jets found
in radio loud quasars \cite{Punsly11} and  is  above the FR~I/FR~II divide at 
$ \bar{Q} \approx  5 \times 10^{43}$erg~s$^{-1}$, suggestive of a 
highly efficient supersonic jet ejection \cite{RS91},  energetic enough to pierce through
the host galaxy's dense environment.   

Therefore, assuming $M_{\rm BH} \sim 2 \times 10^{8} M_{\odot}$, the 
lower of the aforementioned black hole mass estimates for \sp, and assuming a rather strong magnetic field 
$B = 2 \times 10^{4}$~Gauss,  equation~\ref{eq:BZ} above shows that the 
estimated jet power can be achieved via the classic (MHD dominated) BZ process  
for  $a \gtrsim 0.73$, i.e. needing a rapid, yet not a maximally spinning black hole. However, even 
if this black hole is near maximally
spinning ($a =  0.98$), the magnetic field cannot be lower than $ B \approx 1.5 \times 10^{4}$~Gauss, 
highlighting the key role  of magnetic field advection in extracting the spin energy of the hole.
 It is also evident that for the same black hole mass as above,  but for a  very low spin  $a \approx 0.1$, 
the estimated $\bar{Q}$ is not obtained  unless the disk magnetic field   
$ B \gtrsim 1.5 \times 10^{5}$~Gauss. However, such an extremely 
 strong magnetic field might never occur normally,  as the  field strength near the inner accretion disk 
 is not expected to exceed the Eddington  value  $B_{Edd} \approx 6 \times 10^{4} \, M_{8}^{-1/2}$~Gauss
 for accretion at the Eddington rate.  From  mid-IR colors, lack of broad lines  and 
a very weak H${ \alpha}$ flux observed  in the AGN's optical spectrum, we inferred above
 that the accretion state of black hole in \sp  is possibly sub-Eddington.  This is a characteristic 
 of a  radiatively inefficient, low luminosity active 
 galactic nuclei (LLAGN/LINER). Thus, the above analysis suggests that in the launching of highly efficient 
 jetted outflow  in \sp the spin of the black hole and its accretion state might 
be playing  dominant roles,  and additionally, the black hole is  required to be unusually massive,  despite 
being hosted by a spiral galaxy. We point out that  compared to pure BZ process,  
mainly considered here for its  simplicity, 
some more complex hybrid models of jet production, based on a combination of
 Blandford-Payne  \cite{BP82} and Blandford-Znajek  mechanisms \cite{BZ77} have been proposed 
 \citep[e.g.,][]{meier01,NBBS07}  which predict higher jet luminosities under similar physical conditions.   

Is such a rare combination of  unusually high  black hole mass,  spin and 
low accretion state achievable in a bulge-less disk galaxy? 
Recent numerical simulations \cite{Dotti13} of black hole growth (via accretion and mergers) predicts
that at low redshifts the most rapidly rotating black holes ($a > 0.9$) are also  the 
most massive ones ($10^{8} - 10^{9} 
M_{\odot}$), and their spin orientation remain stable over many accretion cycles. However,   such
a black hole  generally ends up in an elliptical galaxy and not in a pure spiral, contrary to the present case.  
For this to happen in a spiral galaxy like \sp  would require 
a coherent, high rate of mass accretion,  whereby  the black hole grows rapidly and  is spun-up via  
matter accreted with a well defined, almost constant angular momentum direction, instead of
major mergers. Due to this unusual 
evolutionary route, the fundamental correlations of the black hole mass  and  galaxy properties 
observed in bulge dominated systems are unlikely to  hold 
in the present case.  Future studies might further explore this important aspect in this and similar systems. 
It is especially  important to verify if the black hole in \sp is indeed more massive than the minimum value
($\sim  10^{8} - 10^{9} M_{\odot}$) inferred in several works for the development of
powerful radio jets in AGN \citep[e.g.,][]{Laor,G-K_M_W,Sikora07,CM11}. Such a large black hole mass will pose a
challenge for the models of growth of  black holes in disk galaxies. Our ideas  
can be better  tested  when in future  
more accurate measurements  of both the mass and spin of  the black hole in \sp become available.

It seems plausible  that the unusual jet activity  of J2345-0449 may have been triggered by 
the AGN accretion state switching 
from  a previous high Eddington rate ($\lambda \gtrsim 10^{-2} - 10^{-3} $)
to a sub-Eddington rate, where 

\begin{equation}
\centering
\lambda = {\dot M} /{{\dot M}_{Edd}}
\label{eq:eddington}
\end{equation}

denotes the dimensionless mass accretion rate in units of  the Eddington accretion rate  ${{\dot M}_{Edd}} $. 
A strong link between jet formation and accretion rate has been discussed in 
the literature \citep[e.g.,][]{Rees82,meier01}, 
which is of direct relevance in the context of the  giant radio jets  found in this unusual spiral galaxy.
It is well known that at accretion rates above a critical value $\lambda_{\rm crit} \sim 10^{-2} - 10^{-3} $ 
the  disk structure  is  a standard Shakura and Sunyaev, 
radiatively efficient, geometrically thin disk \cite{SS73}, whereas at  reduced accretion rates 
($\lambda << \lambda_{\rm crit}  $),    
transition to an advection-dominated accretion flow  (ADAF)  results, fueled by hot gas from 
a large-scale quasi-spherical halo \cite{Narayan94,Narayan95}. In ADAF  most of the 
heat generated by the local, turbulent viscous heating is rapidly advected 
 inside the black hole horizon (and eventually swallowed) or released in a wind or radio-jet.
Thus, the radiative cooling efficiency of the geometrically thick,  optically thin, disk is greatly diminished.
Therefore, one would find black holes in ADAF state to be unusually faint 
(disk bolometric luminosity $L_{bol} << 0.1 {\dot M} c^{2}$); such as the galactic center black hole Sgr A$^*$ 
or the radio galaxy M87/Virgo-A  \cite{Falcke04}. 
In ADAF   the inner region of the accretion disk gets extremely  hot, 
forming a thick ion-torus which helps  to  collimate the
relativistic outflow  into  the `funnel'  regions at the poles \cite{Rees82}. In addition,  
the thick torus  effectively anchors strong poloidal magnetic field  
that is required  to channel the  rotational energy of the hole in the form of   electromagnetic 
fields and particles via the BZ mechanism \cite{BH08}. The magnetic field  plays another important role as 
MHD turbulence in accretion flow is generated by the magnetorotational 
instability  \cite{BH91} which provides the necessary viscous torque for removing 
angular momentum from the accreting gas, thus driving the inflow, and amplifying the large-scale magnetic field.

This is an  attractive  scenario  for   \sp   because the  AGN  in it is presently  
in the   low excitation state (LERG/LLAGN), showing 
extremely low radiative efficiency, but a very 
high kinetic luminosity in  jetted outflows, strongly reminiscent  of an ADAF. This observation provides
an important clue as to how the radio jets in \sp might be launched. 
We further investigate this  rare occurrence  of relativistic jets 
in light of the fundamental paradigm, that, 
inspite of a vast difference in involved black hole mass, length and time scales, almost all relativistic
disk-jet coupled phenomena happen in a scale invariant manner in both radio loud AGNs and the 
galactic `microquasars' \cite{Mirabel}. In galactic black hole X-ray binaries  a transition to `low-hard' X-ray 
state is generally always  accompanied by steady radio jet formation, whereas their radio emission is 
quenched in   `high-soft' X-ray state \cite{FBG}. This indicates a strong link between the jet formation and accretion
state of a black hole. In this context a  `fundamental plane' (FP) of accreting black holes 
which links SMBHs to  stellar mass galactic black holes has been proposed, finding a positive correlation 
between their core radio ($L_{\rm R}$) and X-ray ($L_{\rm X}$) luminosities,  with the black hole mass as a
scaling factor, of the form \cite{Merloni03,Falcke04}

\begin{equation}
\centering
log L_{\rm R}  = 0.60\, log L_{\rm X} + 0.78\, log M_{\rm BH} + 7.33 
\end{equation}

Where $L_{\rm R}$ (at 5~GHz) and  $L_{\rm X}$ (in 2-10 KeV band) are in erg~s$^{-1}$ and the 
black hole mass is in $M_{\odot}$ units. The
FP  is a powerful  scheme which unifies the physics of 
accretion and jet formation for black holes  across an enormous $\sim 10^{1} - 10^{10}\, M_{\odot}$ mass 
range, primarily in their sub-Eddington, low-hard state.  
At present we lack a hard X-ray flux and a robust mass estimate for black hole  in \sp to place it on the 
FP. Nevertheless, assuming central black hole mass is in 
range of $10^{8} \, M_{\odot}$ to $10^{9} \, M_{\odot}$, the above  
FP relation predicts  a core X-ray luminosity
 $L_{\rm X} \sim 2.4 \times 10^{43} $~erg~s$^{-1}$ to $L_{\rm X} \sim 1.2 \times 10^{42}$~erg~s$^{-1}$ 
respectively, using the VLA 5~GHz  core (the jet-base) 
radio luminosity of $L_{\rm R} \sim 4 \times 10^{39} $~erg~s$^{-1}$.
Therefore,  a  future  measurement of $L_{\rm X}$ via X-ray observations, 
and an  independent estimate of the black hole mass would  reveal if the AGN in \sp  
 (putatively in a low-hard ADAF state) conforms to the  FP of accreting black holes or not? Such data 
will provide a powerful means of testing the  disk-jet coupling theories in an extremely unusual 
circumstance, and may also suggest some new, hitherto untested ways of forming relativistic jets. 
However, the fundamental question remains;   why such a transition of accretion state does not 
commonly occur  in other spiral galaxies, 
which almost never show extended radio jets and are mostly radio-quiet?  As we have discussed above, 
the answer could be, that perhaps in our present galaxy the efficiency of jet-launch 
is greatly  boosted  by an unusually massive (and possibly rapidly spinning) SMBH 
(at least a few~$\times 10^{8} M_{\odot}$)  in the galaxy's nucleus, formed via a  predominantly disk 
driven accretion route. Moreover, a strong, almost saturated magnetic field lines anchoring the
accretion disk might have also  enhanced the luminosity of the jet, which can take 
place only via a geometrically thick ADAF \cite{SB13}. Moreover, in the ADAF, unlike the thin-disk case, the
black hole spin is a crucial factor for powerful jet formation \cite{meier01,TNM10,SNPZ13}. 

Such an over-massive, spinning black hole in a pure (spheroidal bulgeless)  
disk galaxy can result only via a previous rapid growth phase  at  high mass accretion rate 
 $\lambda \sim 0.1 - 0.01 $.  In this formative phase, now long past,  the galaxy \sp might have shone like a
luminous quasar  due to large radiative efficiency of a strongly accreting thin  disk. 
In accretion models, the radiative efficiency and growth rate of black hole 
depends only on the dimensionless parameter
 $\lambda$ and not  on the absolute value of $M$  or ${\dot M}$. The e-folding black hole growth time $t_{\rm BH}$ 
 and the black hole mass at a given  time $M{\rm (t)}$ are given by \cite{Shapiro05,HNH06}

\begin{equation} 
\centering
t_{\rm BH} = M/{\dot M} = t_{\rm sal}/\lambda  
\end{equation}
\begin{equation}
\centering
M{\rm (t)} = M{\rm (0)} \exp[\, ((1 - \epsilon)/\epsilon) \, (t/t_{\rm sal}) \, \lambda \,]
\label{eq:BH_grow}
\end{equation}

where, the Salpeter time-scale $t_{\rm sal} = 0.45$~Gyr,   initial  black hole seed mass $M{\rm (0)}$  at $t=0$ and
the radiative efficiency of the accreted mass into energy $\epsilon \approx 0.06 - 0.42$ (for spin $a = 0 - 1$).  
With  gradual mass and angular momentum transfer and a steady increase of the black hole spin $a$, 
the  radiative efficiency $\epsilon$ would also rise (for a radiatively efficient disk),  slowing
down the black hole growth rate. Thus, in the initial phase of black hole growth the mass accretion
rate  needs to be near Eddington ($\lambda \sim 1$) and the spin-up  rate ${\dot a}$  slow  so that  the
black hole may grow to an unusually large final mass via gaseous accretion. We have no estimate for 
the initial  seed mass of the black hole in \sp
 and the cosmic history of its spin and mass evolution is also unknown.
Nevertheless, starting from a Population-III (POP-III) 
seed black hole of $M{\rm (0)} \sim 10^2  M_{\odot} $, formed via stellar route at $z \approx 40$, and  
a coherent, average high Eddington rate accretion ($\bar{\lambda} \approx 0.1$, $\bar{\epsilon} =0.1$), 
it still requires $\sim 18$ e-folding times of 
$t \approx 18 \times t_{\rm sal} \approx 8$~Gyr ($z(t) \approx 0.6$) to grow a 
black hole to $ \sim  10^9  M_{\odot} $ (equation~\ref{eq:BH_grow}). However, this  is shortened  
to  $t \approx 1.8 \times t_{\rm sal} = 0.8$~Gyr (by $z(t) \approx 6.8$) 
for near-Eddington rate accretion ($\bar{\lambda} \approx 1$). For significantly 
lower accretion rates ($\lambda << 1$) 
the growth timescale becomes longer than the age of the universe and the 
black hole cannot possibly accumulate a significant fraction of its mass.
Thus, a relatively short,  initial phase of rapid growth of the  
central black hole at  near Eddington rate accretion  is suggested for \sp, implying that presently we are observing
its AGN  as a faded remnant  of an  erstwhile highly luminous QSO phase.

Therefore,  in our proposed model, the  radio-jet launching spiral galaxy \sp finds itself in an accretion state 
similar to local radio loud ellipticals of low excitation state, in contrast to the
recently discovered  radio loud, narrow-line  Seyfert-1  AGN in spiral galaxies (extreme objects themselves) 
that are inferred to be accreting at near  
Eddington rates ($\lambda \approx 0.1 - 1$), host smaller
mass black holes ($M_{\rm BH} \sim 10^{6} - 10^{8} M_{\odot}$) and are never observed to eject 
$0.1 - 1$~Mpc scale relativistic jets \citep{Doi,Foschini11,Komossa06}. Moreover, existence of this galaxy is  
at odds with the finding that among the AGN population in local 
Universe  essentially all AGN with moderate to large  accretion rates (i.e., $t_{\rm BH} < t_{\rm Hubble}
\sim 14~{\rm Gy}$) are in spirals, whereas their counterparts in almost every elliptical have a negligible 
mass accretion rate ($t_{\rm BH} > t_{\rm Hubble}$) 
\cite{HNH06}.   In \sp an accretion rate-dependent state transition 
 to low radiative efficiency, ADAF state
 might have taken place due to the enormous growth  of its accreting black hole itself and  
 its energetic   feed-back acting on
the accretion flow, heating the latter and thus reducing the AGN fuelling rate 
 below the critical value $\lambda_{\rm crit} \sim 10^{-2} - 10^{-3} $. The two set of radio lobes 
forming the  double-double structure are clearly indicative of an
 intermittent  radio jet activity,  possibly resulting from an   
episodic  switching of the accretion  modes between 
a low (ADAF, powerful radio jets, low radiative efficiency) 
to high (thin disk, no/weak radio jets, high radiative efficiency)  accretion states. Thus, 
finding the black hole's  mass, spin, and the impact of the highly energetic jet outflow on the 
surrounding galaxy is  clearly very important. Perhaps, the most crucial clues for understanding  
this extraordinary  and puzzling radio galaxy would come from its 
future detailed X-ray, UV, infra-red and radio observations. 
This will allow the  complete spectral energy distribution of the 
galaxy and central AGN to be determined for theoretical modelling.
A sensitive X-ray observation of a possible corona of hot, X-ray emitting  
gas around the massive spiral host, a remnant of its formation history,  and detection of an AGN  with hard 
X-ray spectrum (peaking at $\sim 100$~keV) showing low bolometric luminosity  would strengthen 
the model outlined above. The spin of the black hole can be determined by the method of 
continuum fitting the flux and temperature of the X-ray emission from the accretion disk, while the
mass of the black hole can be obtained via high resolution stellar spectroscopy near the black hole's
sphere of influence. High resolution VLBI imaging and polarization mapping of the 
inner radio jets near  the core, in the disk/corona collimation region, would  also be very informative for
understanding the  jet launching process in this highly unusual AGN.

To conclude, the present object \sp is  the most extreme
example of powerful jet activity and  SMBH  growth in a  spiral galaxy, probably arising
via a non-standard evolutionary route that mainly involves internal, merger-free processes
persisting through most of  its formation  history.
Future observations and numerical simulations  should clarify whether its exceptional properties
can be understood within the current frameworks of galaxy formation.

\acknowledgments
On the occasion of 50 years of radio astronomy research at the Tata Institute of Fundamental 
Research (TIFR) and India, as well as 10 years of operation of the Giant Metrewave Radio Telescope (GMRT) 
as an international observatory, we dedicate this work to the vision and  pioneering spirit of 
Professor Govind Swarup, late Professor Vijay Kapahi and many talented Indian scientists, engineers
and technicians who have made it possible. We thank the anonymous referee for providing 
many constructive comments which improved the
paper. We thank  IUCAA Girawali Observatory (IGO)  staff, and
Prof. Vijay Mohan for their  support during  observations. Team members of IFOSC instrument on the IUCAA
2m telescope are thanked for their excellent work. J.J. and BKG acknowledge IUCAA's support under the
Visiting Associate  program. We used archival
data from GMRT, CFHT and VLA facilities.
The GMRT is a national facility
operated by the National Centre for Radio Astrophysics (NCRA) of
the Tata Institute of Fundamental Research, India. Our results are
based on observations obtained with MegaCam, a joint project of CFHT and CEA/DAPNIA, at
the Canada–-France–-Hawaii Telescope,  which is operated by the National Research
Council of Canada, the Institut National des Science de l'Univers of the Centre National de
la Recherche Scientifique of France and the University of Hawaii.
The VLA is a facility of the National Radio Astronomy
Observatory (NRAO). The NRAO is a facility of the National Science
Foundation, operated under cooperative agreement by
Associated Universities, Inc. Funding for SDSS  has been provided by the Alfred P. Sloan Foundation,
the Participating Institutions, the National Science Foundation, and the U.S. Department 
of Energy Office of Science.



\begin{table*}
\begin{minipage}{150mm}
\begin{center}
\caption{{\bf The results of two dimensional bulge-disk decomposition}. Here $m_b$, $m_d$ are the
 magnitudes of bulge and disk components and  $m_{tot}$ is the total magnitude. CFHT$_x$ and SDSS$_x$
   denote the fits  using either CFHT or the SDSS images in  filter band {\em x}. All magnitudes are k-corrected
      and also corrected for galactic extinction, as described in the main text.}
\bigskip
\begin{tabular}{@{}cccc@{}}
\hline
	       Telescope & $m_b$ & $m_d$ & $m_{tot}$ \\
	           (filter) & (mag) & (mag) & (mag)\\
\hline
\hline
			          CFHT$_g$ & $17.14 \pm 0.01$ & $15.18 \pm 0.01$ & $15.02 \pm 0.01$ \\
				             CFHT$_r$ & $16.36 \pm 0.01$ & $14.75 \pm 0.01$ & $14.53 \pm 0.01$  \\
					                     SDSS$_g$ & $17.52 \pm 0.03$ & $15.15 \pm 0.01$ & $15.03 \pm 0.03$ \\
							                           SDSS$_r$ & $16.56 \pm 0.02$ & $14.72 \pm 0.01$ & $14.54 \pm 0.02$ \\
										                                SDSS$_i$ & $16.22 \pm 0.02$ & $14.43 \pm 0.01$ & $14.24 \pm 0.02$  \\
														                                     SDSS$_z$ & $15.72 \pm 0.04$ & $14.34 \pm 0.01$ & $14.07 \pm 0.04$ \\

\end{tabular}
\label{tab:result_mag}
																													      \end{center}
																													      \end{minipage}
																													      \end{table*}

\bigskip
\begin{table*}
\begin{minipage}{150mm}
\begin{center}
\caption{{\bf The results of two dimensional bulge-disk decomposition}. Here n is the S\'ersic index, 
$r_e$, $r_d$ are the bulge and disk scale lengths, and
${b/a}_{b}$, ${b/a}_{d}$ are the axis ratios (semi minor/semi major) of the bulge and disk components respectively.
CFHT$_x$ and SDSS$_x$ denote the fits  using either CFHT or the SDSS images in  filter band {\em x}. The fitting
procedure is described in the main text.
}
\bigskip
\begin{tabular}{@{}ccccccc@{}}
\hline
Telescope & $r_e$ & n & $r_d$ & ${b/a}_b$ & ${b/a}_d$\\
(filter) & (kpc) & & (kpc) & & & \\
 \hline
   \hline
       CFHT$_g$ & $1.18 \pm 0.01$ & $1.05 \pm 0.02$ & $7.51 \pm 0.01$ & $0.56 \pm 0.01$ & $0.53 \pm 0.01$ \\
              CFHT$_r$ & $1.25 \pm 0.01$ & $1.26 \pm 0.02$ & $7.06 \pm 0.01$ & $0.56 \pm 0.01$ & $0.53 \pm 0.01$ \\
			                 SDSS$_g$ & $0.87 \pm 0.13$ & $0.26 \pm 0.58$ & $5.64 \pm 0.03$ & $0.44 \pm 0.08$ & $0.54 \pm 0.01$ \\
					                       SDSS$_r$ & $1.02 \pm 0.07$ & $0.30 \pm 0.26$ & $5.38 \pm 0.03$ & $0.44 \pm 0.04$ & $0.55 \pm 0.01$ \\
							                                    SDSS$_i$ & $1.12 \pm 0.03$ & $0.55 \pm 0.15$ & $5.18 \pm 0.03$ & $0.49 \pm 0.03$ & $0.56 \pm 0.01$ \\
											                                         SDSS$_z$ & $1.09 \pm 0.06$ & $0.95 \pm 0.26$ & $5.10 \pm 0.08$ & $0.66 \pm 0.04$ & $0.54 \pm 0.01$ \\

\end{tabular}
\label{tab:sersic_param}
\end{center}
\end{minipage}
\end{table*}

\begin{table*}
\begin{minipage}{150mm}
\begin{center}
\caption{{\bf  {\em WISE} mid-infrared  magnitudes and luminosities for \sp.} Magnitudes with their errors are
given in the first row in the four mid-IR bands. Second row shows the derived luminosities in erg~s$^{-1}$.
In third row the luminosity is shown in solar units and the fourth row shows the 
signal-to-noise ratios of detection in four bands. } 
\bigskip
\begin{tabular}{@{}ccccccc@{}}
\hline
{\em WISE} band & W1 & W2 & W3 & W4 \\
(Wavelength) & $3.4\,\mu$m &$4.6\,\mu$m & $12\,\mu$m & $22\,\mu$m  \\
 \hline
  \hline
Magnitude & $11.882 \pm 0.023$ & $11.818 \pm 0.023$ & $9.001 \pm 0.028$ & $7.096 \pm 0.100$  \\
Luminosity & $6.71\times10^{43}$ & $2.88\times10^{43}$ & $2.62\times10^{43}$ & $2.26\times10^{43}$  \\
Luminosity ($L_{\odot}$) & $4.04\times10^{11}$ & $4.41\times10^{11}$ & $5.69\times10^{12}$ & $3.35\times10^{13}$  \\
S/N ratio & 47.9 & 47.8 & 38.7 & 10.9  \\		    

\end{tabular}
\label{tab:infrared}
\end{center}
\end{minipage}
\end{table*}

\hfill
\eject


\newpage
\thispagestyle{empty}

\newpage
\clearpage

\newpage

\newpage
\clearpage

\newpage
\clearpage

\newpage
\clearpage
\thispagestyle{empty}

\newpage
\clearpage

\newpage
\clearpage

\newpage
\clearpage


\begin{thebibliography}{}
\expandafter\ifx\csname url\endcsname\relax
\def\url#1{\texttt{#1}}\fi
\expandafter\ifx\csname urlprefix\endcsname\relax\def\urlprefix{URL }\fi
\providecommand{\bibinfo}[2]{#2}
\providecommand{\eprint}[2][]{\url{#2}}

\bibitem[{Abdo et al. 2009}]{abdo09}
\bibinfo{author}{{Abdo}, A.A.}, \bibinfo{author}{{Ackermann}, M.},\bibinfo{author}{{Ajello}, M.}, et al.
\bibinfo{year}{2009},
\newblock \emph{\bibinfo{journal}{\apj}} \textbf{\bibinfo{volume}{707}},
\bibinfo{pages}{L142}

\bibitem[{Ahn et al. 2014}]{Ahn14}
\bibinfo{author}{{Ahn}, C.P.} et al.
\bibinfo{year}{2014},
\newblock \emph{\bibinfo{journal}{\apjs}} \textbf{\bibinfo{volume}{211}},
\bibinfo{pages}{16}


\bibitem[{Akos et al. 2013}]{Akos13}
\bibinfo{author}{{\'Akos}, B.}, \bibinfo{author}{{Forman}, W.R.}, \bibinfo{author}{{Kraft}, R.P.} \& 
\bibinfo{author}{{Jones}, C.},
\bibinfo{year}{2013},
\newblock \emph{\bibinfo{journal}{\apj}} \textbf{\bibinfo{volume}{772}},
\bibinfo{pages}{98}

\bibitem[{Alexander \& Hickox 2012}]{Hickox12}
\bibinfo{author}{{Alexander}, D.M.} \& \bibinfo{author}{{Hickox}, R.C.},
\bibinfo{year}{2012},
\newblock \emph{\bibinfo{journal}{New Astronomy Reviews}} \textbf{\bibinfo{volume}{56}},
\bibinfo{pages}{93}


\bibitem[{Argence \&  Lamareille 2009}]{AL}
\bibinfo{author}{{Argence}, B.} \& \bibinfo{author}{{Lamareille}, F.}
\bibinfo{year}{2009},
\newblock \emph{\bibinfo{journal}{\aap}} \textbf{\bibinfo{volume}{495}},
\bibinfo{pages}{759}



\bibitem[{Balbus \& Hawley 1991}]{BH91}
\bibinfo{author}{{Balbus}, S.A.} \&\bibinfo{author}{{Hawley}, J.F.}
\bibinfo{year}{1991},
\newblock \emph{\bibinfo{journal}{\apj}} \textbf{\bibinfo{volume}{376}},
\bibinfo{pages}{214}

\bibitem[{Baldwin, Phillips \& Terlevich  1981}]{BPT}
\bibinfo{author}{{Baldwin}, J.A.}, \bibinfo{author}{{Phillips}, M.M.} \&  \bibinfo{author}{{Terlevich}, R.}
\bibinfo{year}{1981},
\newblock \emph{\bibinfo{journal}{PASP}} \textbf{\bibinfo{volume}{93}},
\bibinfo{pages}{5}





\bibitem[{Bardeen \& Petterson 1975}]{BP75}
\bibinfo{author}{{Bardeen}, J.M.}\& \bibinfo{author}{{Petterson}, J.A.}
\bibinfo{year}{1975},
\newblock \emph{\bibinfo{journal}{ApJL}} \textbf{\bibinfo{volume}{195}},
\bibinfo{pages}{L65}

\bibitem[{Beckwith \& Hawley 2008}]{BH08}
\bibinfo{author}{{Beckwith}, K.} \&\bibinfo{author}{{Hawley}, J.F.}
\bibinfo{year}{2008},
\newblock \emph{\bibinfo{journal}{\apj}} \textbf{\bibinfo{volume}{678}},
\bibinfo{pages}{1180}

\bibitem[{Begelman et al. 1984}]{BBR}
\bibinfo{author}{{Begelman}, M.C.}, 
\bibinfo{author}{{Blandford}, R.D.}, \bibinfo{author}{{Rees}, M.J.}
\bibinfo{year}{1984},
\newblock \emph{\bibinfo{journal}{Rev. Mod. Phys.}} \textbf{\bibinfo{volume}{56}},
\bibinfo{pages}{255}

\bibitem[{Berti \& Volonteri 2008}]{BV08}
\bibinfo{author}{{Berti}, E.} \& \bibinfo{author}{{Volonteri}, M.}
\bibinfo{year}{2008},
\newblock \emph{\bibinfo{journal}{\apj}} \textbf{\bibinfo{volume}{684}},
\bibinfo{pages}{822} 

\bibitem[{Bertin \& Arnouts 1996}]{ber96}
\bibinfo{author}{{Bertin}, E.} \&  \bibinfo{author}{{Arnouts}, S.}
\bibinfo{year}{1996},
\newblock \emph{\bibinfo{journal}{\aaps}} \textbf{\bibinfo{volume}{117}},
\bibinfo{pages}{393-404}


\bibitem[{Bertin et al. 2002}]{Bertin02}
\bibinfo{author}{{Bertin}, E.}, \bibinfo{author}{{Mellier}, Y.}, \bibinfo{author}{{Radovich}, M.}, 
\bibinfo{author}{{Missonnier}, G.}, \bibinfo{author}{{Didelon}, P.} \& \bibinfo{author}{{Morin}, B.}
\bibinfo{year}{2002},
\newblock \emph{\bibinfo{journal}{Astr. Soc. Pacific Conf. Ser.}}, 
 \textbf{\bibinfo{volume}{281}},
\bibinfo{pages}{228} (Edited by David A. Bohlender, Daniel Durand, and Thomas H. Handley)


\bibitem[{Blandford \& Znajek 1977}]{BZ77}
\bibinfo{author}{{Blandford}, R.D.} \& \bibinfo{author}{{Znajek}, R.L.}
\bibinfo{year}{1977},
\newblock \emph{\bibinfo{journal}{\mnras}} \textbf{\bibinfo{volume}{179}},
\bibinfo{pages}{433}

\bibitem[{Blandford \& Payne 1982}]{BP82}
\bibinfo{author}{{Blandford}, R.D.} \&\bibinfo{author}{{Payne}, D.G.}
\bibinfo{year}{1982},
\newblock \emph{\bibinfo{journal}{\apj}} \textbf{\bibinfo{volume}{199}},
\bibinfo{pages}{883}


\bibitem[{Bouche et al. 2013}]{Bouche13}
\bibinfo{author}{{Bouche}, N.},\bibinfo{author}{{Murphy}, M.T.},\bibinfo{author}{{Kacprzak}, G.G.}, et.al.
\bibinfo{year}{2013},
\newblock \emph{\bibinfo{journal}{Science}} \textbf{\bibinfo{volume}{341}},
\bibinfo{pages}{50}


\bibitem[{Boulade et al. 2013}]{bou03}
\bibinfo{author}{{Boulade}, O.}, et al.
\bibinfo{year}{2003},
\newblock \emph{\bibinfo{journal}{Society of Photo-Optical Instrumentation Engineers (SPIE)
Conference Series}} \textbf{\bibinfo{volume}{4841}},
\bibinfo{pages}{72} 



\bibitem[{Cappellari \& Emsellem 2004}]{Cappellari_Emsellem}
\bibinfo{author}{{Cappellari}, M.} \& \bibinfo{author}{{Emsellem}, E.}
\bibinfo{year}{2004},
\newblock \emph{\bibinfo{journal}{PASP}} \textbf{\bibinfo{volume}{116}},
\bibinfo{pages}{138}


\bibitem[{Chiaberge \& Marconi 2011}]{CM11}
\bibinfo{author}{{Chiaberge}, M.} \& \bibinfo{author}{{Marconi}, A.}
\bibinfo{year}{2011},
\newblock \emph{\bibinfo{journal}{\mnras}} \textbf{\bibinfo{volume}{416}},
\bibinfo{pages}{917}

\bibitem[{Courteau 1997}]{Courteau97}
\bibinfo{author}{{Courteau}, S.}
\bibinfo{year}{1997},
\newblock \emph{\bibinfo{journal}{\aj}} \textbf{\bibinfo{volume}{114}},
\bibinfo{pages}{2402}


\bibitem[{Daly et al. 2012}]{Daly2012}
\bibinfo{author}{{Daly}, R.A.},\bibinfo{author}{{Spinkle}, T.B.},\bibinfo{author}{{O'Dea}, C.P},et.al.
\bibinfo{year}{2012},
\newblock \emph{\bibinfo{journal}{\mnras}} \textbf{\bibinfo{volume}{423}},
\bibinfo{pages}{2498} 


\bibitem[{Dekel et al. 2009}]{Dekel09}
\bibinfo{author}{{Dekel}, A.},\bibinfo{author}{{Birnboim}, Y.},\bibinfo{author}{{Engel}, G.}, et al.
\bibinfo{year}{2009},\newblock \emph{\bibinfo{journal}{Nature}} \textbf{\bibinfo{volume}{457}},
\bibinfo{pages}{451} 



\bibitem[{Danovich et al. 2012}]{Dekel012}

\bibinfo{author}{{Danovich}, M.},\bibinfo{author}{{Dekel}, A.},\bibinfo{author}{{Hahn}, O.}\& \bibinfo{author}{{Teyssier}, R.} 
\bibinfo{year}{2012},
\newblock \emph{\bibinfo{journal}{\mnras}} \textbf{\bibinfo{volume}{422}},
\bibinfo{pages}{1732}

\bibitem[{Doeleman et al. 2012}]{Doeleman12}
\bibinfo{author}{{Doeleman}, S.S.},\bibinfo{author}{{Fish}, V.L.},\bibinfo{author}{{Schenck}, D.E.}, et al.
\bibinfo{year}{2012},
\newblock \emph{\bibinfo{journal}{Science}} \textbf{\bibinfo{volume}{338}},
\bibinfo{pages}{355} 


\bibitem[{Doi 2012}]{Doi}
\bibinfo{author}{{Doi}, A.},\bibinfo{author}{{Nagira}, H.},\bibinfo{author}{{Kawakatu}, N.}, et al.
\bibinfo{year}{2012},
\newblock \emph{\bibinfo{journal}{\apj}} \textbf{\bibinfo{volume}{760}},
\bibinfo{pages}{41} 


\bibitem[{Dotti et al. 2013}]{Dotti13}
\bibinfo{author}{{Dotti}, M.}, \bibinfo{author}{{Colpi}, M.}, \bibinfo{author}{{Pallini}, S.},
\bibinfo{author}{{Perego}, E.} \& \bibinfo{author}{{Volonteri}, M.}
\bibinfo{year}{2013},
\newblock \emph{\bibinfo{journal}{\apj}} \textbf{\bibinfo{volume}{762}},
\bibinfo{pages}{68}



\bibitem[{Dubois et al. 2014}]{Dubois14}
\bibinfo{author}{{Dubois}, Y.}, \bibinfo{author}{{Volonteri}, M.}, \bibinfo{author}{{Silk}, J.},
\bibinfo{author}{{Devriendt}, J.} \& \bibinfo{author}{{Slyz}, A.}
\bibinfo{year}{2014},
\newblock \emph{\bibinfo{journal}{\mnras}} \textbf{\bibinfo{volume}{440}},
\bibinfo{pages}{2333}



\bibitem[{Dunlop et al. 2003}]{Dunlop03}
\bibinfo{author}{{Dunlop}, J.S.}, \bibinfo{author}{{Mclure}, R.J.}, \bibinfo{author}{{Kukula}, M.J.},
\bibinfo{author}{{Baum}, S.A.}, \bibinfo{author}{{O'Dea}, C.P.} \& \bibinfo{author}{{Huges}, D.H.}
\bibinfo{year}{2003},
\newblock \emph{\bibinfo{journal}{\mnras}} \textbf{\bibinfo{volume}{340}},
\bibinfo{pages}{1095} 



\bibitem[{Falcke et al. 2004}]{Falcke04}
\bibinfo{author}{{Falcke}, H.}, \bibinfo{author}{{Kording}, E.} \& \bibinfo{author}{{Markoff}, S.}
\bibinfo{year}{2004},
\newblock \emph{\bibinfo{journal}{\aap}} \textbf{\bibinfo{volume}{414}},
\bibinfo{pages}{895}



\bibitem[{Fanaroff \& Riley 1974}]{FR}
\bibinfo{author}{{Fanaroff}, B.L.}, \& \bibinfo{author}{{Riley}, J.M.}
\bibinfo{year}{1974},
\newblock \emph{\bibinfo{journal}{\mnras}} \textbf{\bibinfo{volume}{167}},
\bibinfo{pages}{31P}



\bibitem[{Fender et al. 2004}]{FBG}
\bibinfo{author}{{Fender}, R.P.},  \bibinfo{author}{{Belloni}, T.M.} \&  \bibinfo{author}{{Gallo}, E.}
\bibinfo{year}{2004},
\newblock \emph{\bibinfo{journal}{\mnras}} \textbf{\bibinfo{volume}{355}},
\bibinfo{pages}{1105}


\bibitem[{Fisher \& Drory 2008}]{Fisher_Drory}
\bibinfo{author}{{Fisher}, D.B.} \& \bibinfo{author}{{Drory}, N.}
\bibinfo{year}{2008},
\newblock \emph{\bibinfo{journal}{\aj}} \textbf{\bibinfo{volume}{136}},
\bibinfo{pages}{773} 


\bibitem[{Foschini 2011}]{Foschini11}
\bibinfo{author}{{Foschini}, L.}
\bibinfo{year}{2011},
\newblock \emph{\bibinfo{journal}{Res. in Astron. Astroph.}} \textbf{\bibinfo{volume}{11}},
\bibinfo{pages}{1266}


\bibitem[{Gadotti 2009}]{Gadotti09}
\bibinfo{author}{{Gadotti}, D.A.}
\bibinfo{year}{2009},
\newblock \emph{\bibinfo{journal}{\mnras}} \textbf{\bibinfo{volume}{393}},
\bibinfo{pages}{1531}


\bibitem[{Gebhardt et al. 2000}]{Gebhardt2000}
\bibinfo{author}{{Gebhardt}, K.}, et al.
\bibinfo{year}{2000},
\newblock \emph{\bibinfo{journal}{\apj}} \textbf{\bibinfo{volume}{539}},
\bibinfo{pages}{L13}



\bibitem[{Giovanelli et al. 1986}]{Giovannelli_86}
\bibinfo{author}{{Giovanelli}, R.},\bibinfo{author}{{Haynes}, M.P.},\bibinfo{author}{{Rubin}, V.C.} \&\bibinfo{author}{{Ford Jr.}, W.C.}
\bibinfo{year}{1986},
\newblock \emph{\bibinfo{journal}{\apj}} \textbf{\bibinfo{volume}{301}},
\bibinfo{pages}{L7} 

\bibitem[{Gopal-Krishna et al. 2008}]{G-K_M_W}
\bibinfo{author}{{Gopal-Krishna}}, \bibinfo{author}{{Mangalam}, A.} \&
\bibinfo{author}{{Wiita}, P.J.}
\bibinfo{year}{2008},
\newblock \emph{\bibinfo{journal}{\apj}} \textbf{\bibinfo{volume}{680}},
\bibinfo{pages}{L13-L16} 


\bibitem[{Governato et al. 2007}]{Governato07}
\bibinfo{author}{{Governato}, F.},\bibinfo{author}{{Willman}, B.},\bibinfo{author}{{Mayer}, L.}, et al.
\bibinfo{year}{2007},
\newblock \emph{\bibinfo{journal}{\mnras}} \textbf{\bibinfo{volume}{374}},
\bibinfo{pages}{1479} 


\bibitem[{Graham 2007}]{Graham07}
\bibinfo{author}{{Graham}, A.W.}
\bibinfo{year}{2007},
\newblock \emph{\bibinfo{journal}{\mnras}} \textbf{\bibinfo{volume}{379}},
\bibinfo{pages}{711}


\bibitem[{Graham et al. 2011}]{Graham}
\bibinfo{author}{{Graham}, A.W.}, \bibinfo{author}{{Onken}, C.A.}, \bibinfo{author}{{Athanassoula}, E.} 
\& \bibinfo{author}{{Combes}, F.}
\bibinfo{year}{2011},
\newblock \emph{\bibinfo{journal}{\mnras}} \textbf{\bibinfo{volume}{412}},
\bibinfo{pages}{2211} 


\bibitem[{Gultekin et al. 2009}]{Gultekin}
\bibinfo{author}{{Gultekin}, K.}, \bibinfo{author}{{Richstone}, D.O.}, \bibinfo{author}{{Gebhardt}, K.}, et al.
\bibinfo{year}{2009},
\newblock \emph{\bibinfo{journal}{\apj}} \textbf{\bibinfo{volume}{698}},
\bibinfo{pages}{198}

\bibitem[{Gurkan et al. 2014}]{GHJ11}
\bibinfo{author}{{Gurkan}, G},
\bibinfo{author}{{Hardcastle}, M.J.} \& \bibinfo{author}{{Jarvis}, M.J.}.
\bibinfo{year}{2014},
\newblock \emph{\bibinfo{journal}{\mnras}} \textbf{\bibinfo{volume}{438}},
\bibinfo{pages}{1149}

\bibitem[{Gwyn 2008}]{gwy08}
\bibinfo{author}{{Gwyn}, S.D.J.}
\bibinfo{year}{2008},
\newblock \emph{\bibinfo{journal}{PASP}} \textbf{\bibinfo{volume}{120}},
\bibinfo{pages}{212} 





\bibitem[{Hawley \& Krolik 2006}]{HK06}
\bibinfo{author}{{Hawley}, J.F.} \& \bibinfo{author}{{Krolik}, J.H.}
\bibinfo{year}{2006},
\newblock \emph{\bibinfo{journal}{\apj}} \textbf{\bibinfo{volume}{641}},
\bibinfo{pages}{103-116}

\bibitem[{Haring \& Rix 2004}]{Haring_Rix}
\bibinfo{author}{{Haring}, N.} \&  \bibinfo{author}{{Rix}, Hans-Walter},
(\bibinfo{year}{2004}),
\newblock \emph{\bibinfo{journal}{\apj}} \textbf{\bibinfo{volume}{604}},
\bibinfo{pages}{L89} 

\bibitem[{Hopkins et al. 2006}]{HNH06}
\bibinfo{author}{{Hopkins}, P.F.}, \bibinfo{author}{{Narayan}, R.} \&\bibinfo{author}{{Hernquist}, L.}
\bibinfo{year}{2006},
\newblock \emph{\bibinfo{journal}{\apj}} \textbf{\bibinfo{volume}{643}},
\bibinfo{pages}{641}

\bibitem[{Hota et al. 2011}]{Hota_2011}
\bibinfo{author}{{Hota}, A.},\bibinfo{author}{{Sirothia}, S.K.},\bibinfo{author}{{Ohyama}, Y.}, et al.
\bibinfo{year}{2011},
\newblock \emph{\bibinfo{journal}{\mnras}} \textbf{\bibinfo{volume}{417}},
\bibinfo{pages}{L36} 

\bibitem[{Hu 2008}]{Hu}
\bibinfo{author}{{Hu}, Jian}
\bibinfo{year}{2008},
\newblock \emph{\bibinfo{journal}{\mnras}} \textbf{\bibinfo{volume}{386}},
\bibinfo{pages}{2242} 



\bibitem[{2009}]{Jones09}
\bibinfo{author}{{Jones}, D.}, et al.
\bibinfo{year}{2009},
\newblock \emph{\bibinfo{journal}{\mnras}} \textbf{\bibinfo{volume}{399}},
\bibinfo{pages}{683}




\bibitem[{Kennicut 1992}]{Kennicut}
\bibinfo{author}{{Kennicut}, R.C. (Jr.)}
\bibinfo{year}{1992},
\newblock \emph{\bibinfo{journal}{\apj}} \textbf{\bibinfo{volume}{388}},
\bibinfo{pages}{310}



\bibitem[{Keres et al. 2005}]{Keres05}
\bibinfo{author}{{Keres}, D.}, \bibinfo{author}{{Katz}, N.}, \bibinfo{author}{{Weinberg}, D.H}, \&
\bibinfo{author}{{Dav\'e}, R.}
\bibinfo{year}{2005},
\newblock \emph{\bibinfo{journal}{\mnras}} \textbf{\bibinfo{volume}{362}},
\bibinfo{pages}{2}

\bibitem[{Komossa et al. 2006}]{Komossa06}
\bibinfo{author}{{Komossa}, S.}, et al
\bibinfo{year}{2006},
\newblock \emph{\bibinfo{journal}{\aj}} \textbf{\bibinfo{volume}{132}},
\bibinfo{pages}{531}


\bibitem[{Kormendy \& Rishstone 1995}]{KR95}
\bibinfo{author}{{Kormendy}, J.} \& \bibinfo{author}{{Richstone}, D.}
\bibinfo{year}{1995},
\newblock \emph{\bibinfo{journal}{Ann. Rev. Astron. Astroph.}} \textbf{\bibinfo{volume}{33}},
\bibinfo{pages}{581}




\bibitem[{Kormendy et al. 2010}]{Kormendy_etal}
\bibinfo{author}{{Kormendy}, J.}, \bibinfo{author}{{Drory}, N.}, \bibinfo{author}{{Bender}, R.}, \& 
\bibinfo{author}{{Cornell}, M.E.}
\bibinfo{year}{2010},
\newblock \emph{\bibinfo{journal}{\apj}} \textbf{\bibinfo{volume}{723}},
\bibinfo{pages}{54} 

\bibitem[{Kormendy et al. 2011}]{Kormendy_nature}
\bibinfo{author}{{Kormendy}, J.}, \bibinfo{author}{{Bender}, R.} \&
\bibinfo{author}{{Cornell}, M.E.}
\bibinfo{year}{2011},
\newblock \emph{\bibinfo{journal}{Nature}} \textbf{\bibinfo{volume}{469}},
\bibinfo{pages}{374} 

\bibitem[{Laor 2000}]{Laor}
\bibinfo{author}{{Laor}, A.}
\bibinfo{year}{2000},
\newblock \emph{\bibinfo{journal}{\apj}} \textbf{\bibinfo{volume}{543}},
\bibinfo{pages}{L111-L114} 

\bibitem[{Ledlow et al. 1998}]{Ledlow_Owen_keel}
\bibinfo{author}{{Ledlow}, M.J.},
\bibinfo{author}{{Owen}, F.N.} \& \bibinfo{author}{{Keel}, W.C.}.
\bibinfo{year}{1998},
\newblock \emph{\bibinfo{journal}{\apj}} \textbf{\bibinfo{volume}{495}},
\bibinfo{pages}{227} 

\bibitem[{Lequeux 1983}]{Lequeux}
\bibinfo{author}{{Lequeux}, J.}
\bibinfo{year}{1983},
\newblock \emph{\bibinfo{journal}{\aap}} \textbf{\bibinfo{volume}{125}},
\bibinfo{pages}{394} 


\bibitem[{Lynden-Bell 1969}]{Lynden_Bell69}
\bibinfo{author}{{Lynden-Bell}, D.}
\bibinfo{year}{1969},
\newblock \emph{\bibinfo{journal}{Nature}} \textbf{\bibinfo{volume}{223}},
\bibinfo{pages}{690} 

\bibitem[{MacDonald \& Thorne 1982}]{MT82}
\bibinfo{author}{{MacDonald}, D.} \& \bibinfo{author}{{Thorne}, K.S.}
\bibinfo{year}{1982},
\newblock \emph{\bibinfo{journal}{\mnras}} \textbf{\bibinfo{volume}{198}},
\bibinfo{pages}{345}


\bibitem[{2007}]{Machalski07}
\bibinfo{author}{{Machalski}, J.},  \bibinfo{author}{{Koziel-Wierzbowska}, D.} \&  
\bibinfo{author}{{Jamrozy}, M.}
\bibinfo{year}{2007},
\newblock \emph{\bibinfo{journal}{Acta Astronomica}} \textbf{\bibinfo{volume}{57}},
\bibinfo{pages}{227}


\bibitem[{Magorrian et al. 1998}]{Magorrian98}
\bibinfo{author}{{Magorrian}, J.} et al.
\bibinfo{year}{1998},
\newblock \emph{\bibinfo{journal}{\apj}} \textbf{\bibinfo{volume}{115}},
\bibinfo{pages}{2285}

\bibitem[{Marconi \& Hunt 2003}]{Marconi_Hunt}
\bibinfo{author}{{Marconi}, A.} \&  \bibinfo{author}{{Hunt}, L.K.},
\bibinfo{year}{2003},
\newblock \emph{\bibinfo{journal}{\apj}} \textbf{\bibinfo{volume}{589}},
\bibinfo{pages}{L21} 


\bibitem[{Martin et al. 2005}]{martin05}
\bibinfo{author}{{Martin}, D.C.} et al.
\bibinfo{year}{2005},
\newblock \emph{\bibinfo{journal}{\apj}} \textbf{\bibinfo{volume}{619}},
\bibinfo{pages}{L1}





\bibitem[{McGaugh 2005}]{McGaugh05}
\bibinfo{author}{{McGaugh}, S.S.} 
\bibinfo{year}{2005},
\newblock \emph{\bibinfo{journal}{\apj}} \textbf{\bibinfo{volume}{632}},
\bibinfo{pages}{859}




\bibitem[{Meier 2001}]{meier01}
\bibinfo{author}{{Meier}, D.L.} 
\bibinfo{year}{2001},
\newblock \emph{\bibinfo{journal}{\apj}} \textbf{\bibinfo{volume}{548}},
\bibinfo{pages}{L9}

\bibitem[{Merloni et al. 2003}]{Merloni03}
\bibinfo{author}{{Merloni}, A.}, \bibinfo{author}{{Heinz}, S.} \& \bibinfo{author}{{De Matteo}, T.}
\bibinfo{year}{2003},
\newblock \emph{\bibinfo{journal}{\mnras}} \textbf{\bibinfo{volume}{345}},
\bibinfo{pages}{1057}


\bibitem[{Mirabel \& Rodriguez 1994}]{Mirabel}
\bibinfo{author}{{Mirabel}, F.} \& \bibinfo{author}{{Rodriguez}, L.F.}
\bibinfo{year}{1994},
\newblock \emph{\bibinfo{journal}{Nature}} \textbf{\bibinfo{volume}{371}},
\bibinfo{pages}{46}


\bibitem[{Morganti 2011}]{Morganti}
\bibinfo{author}{{Morganti}, R.} et al.
\bibinfo{year}{2011},
\newblock \emph{\bibinfo{journal}{\aap}} \textbf{\bibinfo{volume}{535}},
\bibinfo{pages}{A97} 



\bibitem[{Narayan \& Yi 1994}]{Narayan94}
\bibinfo{author}{{Narayan}, R.} \&\bibinfo{author}{{Yi}, Insu}
\bibinfo{year}{1994},
\newblock \emph{\bibinfo{journal}{\apj}} \textbf{\bibinfo{volume}{428}},
\bibinfo{pages}{L13}

\bibitem[{Narayan \& Yi 1995}]{Narayan95}
\bibinfo{author}{{Narayan}, R.} \&\bibinfo{author}{{Yi}, Insu}
\bibinfo{year}{1995},
\newblock \emph{\bibinfo{journal}{\apj}} \textbf{\bibinfo{volume}{452}},
\bibinfo{pages}{710}


\bibitem[{Nemmen et al. 2007}]{NBBS07}
\bibinfo{author}{{Nemmen}, R. S.}, \bibinfo{author}{{Bower}, R. G.}, \bibinfo{author}{{Babul}, A.} \&
\bibinfo{author}{{Storchi-Bergmann}, T.}
\bibinfo{year}{2007},
\newblock \emph{\bibinfo{journal}{\mnras}} \textbf{\bibinfo{volume}{377}},
\bibinfo{pages}{1652}




\bibitem[{OcVirk et al. 2008}]{OcVirk08}
\bibinfo{author}{{OcVirk}, P.}, \bibinfo{author}{{Pichon}, C.},  \&
\bibinfo{author}{{Teyssier}, R.}
\bibinfo{year}{2008}
\newblock \emph{\bibinfo{journal}{\mnras}} \textbf{\bibinfo{volume}{390}},
\bibinfo{pages}{1326}



\bibitem[{Peng et al. 2002}]{pen02}
\bibinfo{author}{{Peng}, C.~Y.}, \bibinfo{author}{{Ho}, L.~C.}, \bibinfo{author}{{Impey}, C.~D.} \&
\bibinfo{author}{{Rix}, H.-W.}
\bibinfo{year}{2002}
\newblock \emph{\bibinfo{journal}{\aj}} \textbf{\bibinfo{volume}{124}},
\bibinfo{pages}{266}



\bibitem[{Penna et al. 2013}]{PNS13}
\bibinfo{author}{{Penna}, R.F.}, \bibinfo{author}{{Narayan}, R.} \& \bibinfo{author}{{Sadowski}, A.}
\bibinfo{year}{2013},
\newblock \emph{\bibinfo{journal}{\mnras}} \textbf{\bibinfo{volume}{436}},
\bibinfo{pages}{3741}





\bibitem[{Pettini \& Pagel 2004}]{Petini&Pagel}
\bibinfo{author}{{Pettini}, M.} \& \bibinfo{author}{{Pagel}, B.E.J.}\ 
\bibinfo{year}{2004},
\newblock \emph{\bibinfo{journal}{\mnras}} \textbf{\bibinfo{volume}{348}},
\bibinfo{pages}{L59}



\bibitem[{Poggianti 1997}]{pog97}
\bibinfo{author}{{Poggianti}, B.~M.}
\bibinfo{year}{1997},
\newblock \emph{\bibinfo{journal}{\aaps}} \textbf{\bibinfo{volume}{122}},
\bibinfo{pages}{399-407}



\bibitem[{Punsly \& Zhang 2011}]{Punsly11}
\bibinfo{author}{{Punsly}, B.} \& \bibinfo{author}{{Zhang}, H.}
\bibinfo{year}{2011},
\newblock \emph{\bibinfo{journal}{\apj}} \textbf{\bibinfo{volume}{735}},
\bibinfo{pages}{L3}




\bibitem[{Rawlings \& Saunders 1991}]{RS91}
\bibinfo{author}{{Rawlings}, S.} \& \bibinfo{author}{{Saunders}, R.}
\bibinfo{year}{1991},
\newblock \emph{\bibinfo{journal}{Nature}} \textbf{\bibinfo{volume}{349}},
\bibinfo{pages}{138}

\bibitem[{Rees et al. 1982}]{Rees82}
\bibinfo{author}{{Rees}, M.J.}, \bibinfo{{Begelman}, M.C.},  
\bibinfo{author}{{Blandford}, R.D.} \& \bibinfo{author}{{Phinney}, E.S.}
\bibinfo{year}{1982},
\newblock \emph{\bibinfo{journal}{Nature}} \textbf{\bibinfo{volume}{295}},
\bibinfo{pages}{17}

\bibitem[{Rigopoulou et al. 2002}]{Rigopoulou_2002}
\bibinfo{author}{{Rigopoulou}, D.},\bibinfo{author}{{Franceschini}, A.},\bibinfo{author}{{Aussel}, H.},et.al.
\bibinfo{year}{2002},
\newblock \emph{\bibinfo{journal}{\apj}} \textbf{\bibinfo{volume}{580}},
\bibinfo{pages}{789} 


\bibitem[{Sadowski et al. 2013}]{SNPZ13}
\bibinfo{author}{{Sadowski}, A.},  \bibinfo{author}{{Narayan}, R.},  \bibinfo{author}{{Penna}, R.F.} \&
\bibinfo{author}{{Zhu}, Y.}
\bibinfo{year}{2013},
\newblock \emph{\bibinfo{journal}{\mnras}} \textbf{\bibinfo{volume}{436}},
\bibinfo{pages}{3856}



\bibitem[{Saikia \& Jamrozy 2009}]{Saikia_Jamrozy_2009}
\bibinfo{author}{{Saikia}, D.J.}, \& \bibinfo{author}{{Jamrozy}, M.}
\bibinfo{year}{2009},
\newblock \emph{\bibinfo{journal}{Bull. Astron. Soc. India}} \textbf{\bibinfo{volume}{37}},
\bibinfo{pages}{63} 


\bibitem[{Schlegel et al. 1998}]{sch98}
\bibinfo{author}{{Schlegel}, D.~J.}, \bibinfo{author}{{Finkbeiner}, D.~P.} \& \bibinfo{author}{{Davis}, M.}
\bibinfo{year}{1998},
\newblock \emph{\bibinfo{journal}{\apj}} \textbf{\bibinfo{volume}{500}},
\bibinfo{pages}{525}




\bibitem[{Schoenmakers 2000}]{Schoenmakers_2000}
\bibinfo{author}{{Schoenmakers}, A.P.},\bibinfo{author}{{de Bruyn}, A.G.},\bibinfo{author}{{Rottgering}, H.J.A.},
\bibinfo{author}{{van der Laan}, H.},\& \bibinfo{author}{{Kaiser}, C.R.}
\bibinfo{year}{2000},
\newblock \emph{\bibinfo{journal}{\mnras}} \textbf{\bibinfo{volume}{315}},
\bibinfo{pages}{371} 


\bibitem[{Schwope et al. 2000}]{Schwope}
\bibinfo{author}{{Schwope}, A.} et al.
\bibinfo{year}{2000},
\newblock \emph{\bibinfo{journal}{Astronomische Nachrichten}} \textbf{\bibinfo{volume}{321}},
\bibinfo{pages}{1}



\bibitem[{Shakura \& Sunyaev 1973}]{SS73}
\bibinfo{author}{{Shakura}, N.L.} \& \bibinfo{author}{{Sunyaev}, R.A.}
\bibinfo{year}{1973},
\newblock \emph{\bibinfo{journal}{\aap}} \textbf{\bibinfo{volume}{24}},
\bibinfo{pages}{337}



\bibitem[{Shankar et al. 2012}]{shankar2012}
\bibinfo{author}{{Shankar}, F.}, \bibinfo{author}{{Marulli}, F.}, \bibinfo{author}{{Mathur}, S.},
\bibinfo{author}{{Bernardi}, M.} \& \bibinfo{author}{{Bournaud}, F.}
\bibinfo{year}{2012},
\newblock \emph{\bibinfo{journal}{\aap}} \textbf{\bibinfo{volume}{540}},
\bibinfo{pages}{A23} 

\bibitem[{Shapiro 2005}]{Shapiro05}
\bibinfo{author}{{Shapiro}, S.L.}
\bibinfo{year}{2005},
\newblock \emph{\bibinfo{journal}{\apj}} \textbf{\bibinfo{volume}{620}},
\bibinfo{pages}{59-68} 

\bibitem[{Sikora et al. 2007}]{Sikora07}
\bibinfo{author}{{Sikora}, M.}, \bibinfo{author}{{Stawarz}, L.} \& \bibinfo{author}{{Lasota}, Jean-Pierre}
\bibinfo{year}{2007},
\newblock \emph{\bibinfo{journal}{\apj}} \textbf{\bibinfo{volume}{658}},
\bibinfo{pages}{815} 

\bibitem[{Sikora \& Begelman 2013}]{SB13}
\bibinfo{author}{{Sikora}, M.} \& \bibinfo{author}{{Begelman}, M.C.} 
\bibinfo{year}{2013},
\newblock \emph{\bibinfo{journal}{\apj}} \textbf{\bibinfo{volume}{764}},
\bibinfo{pages}{L24}

\bibitem[{Simmons et al. 2013}]{Simmons13}
\bibinfo{author}{{Simmons}, B.D.},\bibinfo{author}{{Lintott}, C.},\bibinfo{author}{{Schawinski}, K.}, et al.
\bibinfo{year}{2013},
\newblock \emph{\bibinfo{journal}{\mnras}} \textbf{\bibinfo{volume}{429}},
\bibinfo{pages}{2199}


\bibitem[{Sofue \& Rubin 2001}]{Sofue_Rubin}
\bibinfo{author}{{Sofue}, Y.} \& \bibinfo{author}{{Rubin}, V.C.} 
\bibinfo{year}{2001},
\newblock \emph{\bibinfo{journal}{Ann. Rev. Astro. Astrophys.}} \textbf{\bibinfo{volume}{39}}, 
\bibinfo{pages}{137}





\bibitem[{Soltan 1982}]{Soltan82}
\bibinfo{author}{{Soltan}, A.}
\bibinfo{year}{1982},
\newblock \emph{\bibinfo{journal}{\mnras}} \textbf{\bibinfo{volume}{200}},
\bibinfo{pages}{115}



\bibitem[{Stern et al. 2012}]{Stern12}
\bibinfo{author}{{Stern}, D.}, et al.
\bibinfo{year}{2012},
\newblock \emph{\bibinfo{journal}{\apj}} \textbf{\bibinfo{volume}{753}},
\bibinfo{pages}{30}



\bibitem[{Tchekhovskoy et al. 2010}]{TNM10}
\bibinfo{author}{{Tchekhovskoy}, A.}, \bibinfo{author}{{Narayan}, R.} \& \bibinfo{author}{{McKinney}, J.C.}
\bibinfo{year}{2010},
\newblock \emph{\bibinfo{journal}{\apj}} \textbf{\bibinfo{volume}{711}},
\bibinfo{pages}{50} 

\bibitem[{Tchekhovskoy et al. 2011}]{TNM11}
\bibinfo{author}{{Tchekhovskoy}, A.}, \bibinfo{author}{{Narayan}, R.} \& \bibinfo{author}{{McKinney}, J.C.}
\bibinfo{year}{2011},
\newblock \emph{\bibinfo{journal}{\mnras}} \textbf{\bibinfo{volume}{418}},
\bibinfo{pages}{L79} 

\bibitem[{Thorne 1974}]{Thorne74}
\bibinfo{author}{{Thorne}, K.S.}
\bibinfo{year}{1974},
\newblock \emph{\bibinfo{journal}{\apj}} \textbf{\bibinfo{volume}{191}},
\bibinfo{pages}{507}


\bibitem[{Van den Bosch et al. 2012}]{van-Der-Bosch_Nature}
\bibinfo{author}{{van den Bosch}, R.C.E.},\bibinfo{author}{{Gebhardt}, K.}, \bibinfo{author}{{Gultekin}, K.},et al.
\bibinfo{year}{2012},
\newblock \emph{\bibinfo{journal}{\nat}} \textbf{\bibinfo{volume}{491}},
\bibinfo{pages}{729} 


\bibitem[{Vazdekis 1999}]{Vazdekis99}
\bibinfo{author}{{Vazdekis}, A.}
\bibinfo{year}{1999},
\newblock \emph{\bibinfo{journal}{\apj}} \textbf{\bibinfo{volume}{513}},
\bibinfo{pages}{224}



\bibitem[{Vikram et al. 2010}]{Vinu10}
\bibinfo{author}{{Vikram}, V.},\bibinfo{author}{{Wadadekar}, Y.}, \bibinfo{author}{{Kembhavi}, A.K.},
\bibinfo{author}{{Vijayagovindan}, G.V}
\bibinfo{year}{2010},
\newblock \emph{\bibinfo{journal}{\mnras}} \textbf{\bibinfo{volume}{409}},
\bibinfo{pages}{1379}


\bibitem[{Voges et al. 1999}]{RASS}
\bibinfo{author}{{Voges}, W.} et al.
\bibinfo{year}{1999},
\newblock \emph{\bibinfo{journal}{\aap}} \textbf{\bibinfo{volume}{349}},
\bibinfo{pages}{389}


\bibitem[{Wang et al. 2006}]{Wang06}
\bibinfo{author}{{Wang}, Jian-Min}, \bibinfo{author}{{Chen}, Yan-Mei}, \bibinfo{author}{{Luis}, C.Ho}\&
\bibinfo{author}{{Ross}, J.M.}
\bibinfo{year}{2006},
\newblock \emph{\bibinfo{journal}{\apj}} \textbf{\bibinfo{volume}{642}},
\bibinfo{pages}{L111-L114} 


  \bibitem[{Wilson \& Colbert 1995}]{WC95}
  \bibinfo{author}{{Wilson}, A.S.} \& \bibinfo{author}{{Colbert}, E.J.M.}
  \bibinfo{year}{1995},
  \newblock \emph{\bibinfo{journal}{\apj}} \textbf{\bibinfo{volume}{438}},
  \bibinfo{pages}{62} 

  \bibitem[{Wright et al. 2010}]{Wright10}
  \bibinfo{author}{{Wright}, E.L.},\bibinfo{author}{{Eisenhardt}, Peter R.M.},\bibinfo{author}{{Mainzer}, A.K.}, et al.
  \bibinfo{year}{2010},
  \newblock \emph{\bibinfo{journal}{\aj}} \textbf{\bibinfo{volume}{140}},
  \bibinfo{pages}{1868}




\end{thebibliography}
\end{document}